\documentclass[a4paper,11pt]{article}
\pdfoutput=1
\usepackage{jheppub}
\usepackage[utf8]{inputenc}
\usepackage{amsmath} 
\usepackage{amssymb,amsfonts}
\usepackage{stmaryrd}
\usepackage[all]{xy}
\usepackage{graphicx}
\usepackage{lscape}
\usepackage{tikz}
\usepackage{tikz-3dplot}
\usepackage{float}
\usetikzlibrary{arrows}
\usepackage{rotating}
\usepackage{url}
\usepackage{hyperref}
\usepackage{afterpage}

\usepackage{accents}

\newcommand{\bl}{\bar{\lambda}}
\newcommand{\lr}{\lambda^{\rho}}

\tikzstyle{styleofbox} = [rectangle, rounded corners, minimum width=1cm, minimum height=.5cm,text centered, draw=black, fill={rgb:orange,1;yellow,2;pink,5}]
\tikzstyle{arrow} = [thin,->,>=angle 60]
\tikzstyle{styleofbox2} = [rectangle, rounded corners, minimum width=1cm, minimum height=.5cm,text centered, draw=black, fill={rgb:orange,0;yellow,2;pink,100}]
\tikzstyle{emptybox} = [rectangle, rounded corners, minimum width=.5cm, minimum height=.5cm,text centered, draw=white, fill={white}]
\tikzstyle{styleofbox3} = [rectangle, rounded corners, minimum width=1cm, minimum height=.5cm,text centered, draw=black, fill={rgb:orange,3;yellow,6;pink,10}]
\tikzstyle{styleofbox4} = [rectangle, rounded corners, minimum width=1cm, minimum height=.5cm,text centered, draw=black, fill={rgb:orange,0;yellow,5;pink,50}]
\tikzstyle{styleofbox5} = [rectangle, rounded corners, minimum width=1cm, minimum height=.5cm,text centered, draw=black, fill={rgb:orange,20;yellow,5;pink,50}]
\tikzstyle{styleofbox6} = [rectangle, rounded corners, minimum width=1cm, minimum height=.5cm,text centered, draw=black, fill={rgb:orange,20;yellow,5;pink,0}]
\tikzset{
roundnode/.style={circle, draw, very thick, minimum size=.5mm},
}
\usetikzlibrary{patterns,snakes}

\title{\boldmath The Conformal Characters}

\author[a]{Antoine Bourget}
\author[b]{and Jan Troost}

\affiliation[a]{Department of Physics, Universidad de Oviedo,\\ Calle Federico García Lorca, 8, 33007 Oviedo, Spain}
\emailAdd{bourgetantoine@uniovi.es}
\affiliation[b]{D\'epartement de Physique, \'Ecole Normale Sup\'erieure, \\ CNRS, PSL Research University, Sorbonne Universit\'es, \\ Paris, France}
\emailAdd{troost@lpt.ens.fr}

\abstract{ We revisit the study of the multiplets of the conformal algebra in any dimension. The theory of highest weight  representations is reviewed in the context of the Bernstein-Gelfand-Gelfand category of modules. The Kazhdan-Lusztig polynomials  code the relation between the Verma modules and the irreducible modules in the category and are the key to the characters of the conformal multiplets (whether finite dimensional, infinite dimensional, unitary or non-unitary). We discuss the representation theory  and review in full generality which representations are unitarizable. The mathematical theory that allows for both the general treatment of characters and the full analysis of unitarity is made accessible. A good understanding of the mathematics of conformal multiplets renders the treatment of all highest weight representations in any dimension uniform, and  provides an overarching comprehension of case-by-case results. Unitary highest weight representations and their characters are classified and computed in terms of data associated to cosets of the Weyl group of the conformal algebra.
An executive summary is provided, as well as look-up tables up to and including rank four.
}

\begin{document} 
\maketitle
\flushbottom

\section{Introduction}
Quantum field theory is one of the most successful tools of theoretical physics. It is ubiquitous in our understanding of physical phenomena from the smallest to the largest scales. Conformal quantum field theories can be viewed as simplified quantum field theories that arise at very low or very high energy,
or at critical points. Their symmetry algebra is enlarged. Relativistic conformal field theories allow for a symmetry algebra which includes the conformal algebra $\mathfrak{so}(2,n)$ in  space-time dimension $(n-1)+1$.

It is useful to gather the spectrum of a physical theory in terms of multiplets of the symmetry algebra. Hence it is crucial to study the representation theory of the conformal algebra $\mathfrak{so}(2,n)$. In physical theories, often only highest weight representations will arise. Moreover, in unitary theories, these representations are required to be unitarizable. Thus, the study of the unitary highest weight representations of $\mathfrak{so}(2,n)$ has been an integral part of the physics literature of the last fifty years. 

Importantly, before physicists classified conformal multiplets in all cases of their interest, the mathematics
literature yielded an overarching insight into the generic case, providing a complete classification of conformal multiplets,
with proof. 
In particular, the representation theory will reduce to a theory of Weyl groups, and numbers associated
to pairs of Weyl group elements. That provides an efficient coding of otherwise lengthy manipulations of conformal algebra generators.
In most cases, the mathematics literature precedes the physics literature, which is indicative of the fact that physicists have found the mathematics literature hard to read. We intend to bridge this unfortunate gap in the treatment of this central problem in quantum field theory by providing a physicist's guide to the mathematics literature. Our treatment will be practical yet generic, referring to the relevant mathematics books and papers for the complete proofs while still providing the conformal field theorist with all the necessary tools to reconstruct a particular result using general principles only.

Bridges between the mathematics literature and the physics literature have been constructed previously.
We refer e.g. to \cite{Ferrara:2000nu} for the exploitation of the generic classification of unitary multiplets, and to \cite{Penedones:2015aga} for a review of the salient properties of parabolic
Kazhdan-Lusztig polynomials relevant to conformal multiplets finitely represented on the compact
subalgebra of the conformal algebra. We will provide a considerably more complete treatment, and hopefully a more accessible bridge.
From the effort we invested in identifying and marrying the mathematics and physics literature, we concluded that an introduction for physicists to the intricate mathematics of conformal multiplets remained overdue. 
It is possible to extend the scope of this work to include representations of superalgebras, but in that situation additional subtleties arise -- in particular, the Weyl group geometry alone does not determine completely the representation theory --  we plan to discuss this in future work.

The structure of the paper is as follows. 
In section \ref{warmup}, we treat a warm-up example, namely the $\mathfrak{so}(2,3)$ conformal algebra in three space-time
dimensions. We compute the characters of all irreducible highest weight multiplets, and gently introduce some of the mathematics necessary to understand the structure of the representation theory. We also identify all  multiplets that are unitarizable, and write out their characters in both a mathematical and physical language. After the warm-up section, we introduce the advanced mathematics to treat the
generic case.
In section \ref{characters}, we summarize how to compute characters of all highest weight representations of the algebras $\mathfrak{so}(2,n)$
with $n$ arbitrary. This discussion will include finite dimensional, infinite dimensional, unitary and non-unitary representations. We review how the multiplicities of the irreducible modules in the Verma modules are given by the evaluation of Kazhdan-Lusztig polynomials at argument equal to one, and how the inversion of the decomposition fixes the irreducible characters. It will be sufficient to do calculations in the Weyl group (and Hecke algebra) of the conformal algebra to understand the full structure of the representation theory. In order to apply the formulas, we gather data on the Weyl groups of the $B_k$ and $D_k$ algebras, and the corresponding Kazhdan-Lusztig polynomials.

In section \ref{unitary}, we review how to identify the unitarizable representations among all those studied in section \ref{characters}. We will implicitly make use of necessary and sufficient inequalities
on the quantum numbers which are elegantly derived in the mathematics literature.
In section \ref{factorization} we exploit the specific features of unitary representations to simplify the generic Kazhdan-Lusztig theory, and factorize a compact subalgebra Weyl group, which leads to the study of parabolic Kazhdan-Lusztig theory. That allows us to compute all unitary highest weight 
characters in section \ref{allunitaries}. 

Hasty readers can jump directly to section \ref{physics} where they will find an executive summary, with references to an appendix
containing low rank unitary character tables.

\section{\texorpdfstring{ Warm-up : The $\mathfrak{so}(2,3)$ Algebra}{}}
\label{warmup}

In this section, before facing the representation theory of the $\mathfrak{so}(2,n)$ algebras in all its complexity, we  focus on the conformal algebra $\mathfrak{so}(2,3)$. 
We review the conformal multiplets which have a highest weight. We  determine the
structure of the irreducible representations, and also  which irreducible highest weight
representations are unitarizable. Our analysis is phrased in the mathematical language of the category
of highest weight modules, and introduces a number of useful mathematical concepts. These serve
as a warm-up for the introduction of  more advanced concepts in section \ref{characters}. References for proofs of the statements in this section are mostly postponed to  section \ref{characters} as well. 

\subsection{\texorpdfstring{The Representations of the $\mathfrak{so}(3)$ Algebra}{}}
We draw inspiration from the highest weight representation theory of the simplest Lie algebra $\mathfrak{so}(3)=\mathfrak{su}(2)$, generated by three generators, $\mathfrak{so}(3)= \langle  J_1 , J_2 , J_3 \rangle$. Its representation theory is obtained by first choosing a Cartan subalgebra $\mathfrak{h} = \langle  J_3 \rangle$, as well as raising and lowering operators $J_{\pm} = J_1 \pm i J_2$. Then, we pick a highest weight eigenvector of the Cartan generator $J_3$ with eigenvalue $\lambda$, called the highest weight. The highest weight vector is by definition annihilated by the raising operator. We then act on it with the lowering operator, generating new vectors, which generate a representation of $\mathfrak{so}(3)$. As is well-known, if $\lambda \notin \mathbb{Z}_{\geq 0}$ (in a given normalization), the process never stops and the representation is infinite dimensional. On the other hand, when $\lambda \in \mathbb{Z}_{\geq 0}$, we can consistently define a $\lambda +1$ dimensional (irreducible) representation, with a lowest weight vector which is annihilated by the lowering operator. In this subsection, we formalize these well-known facts in the language of modules, which is used in the rest of the paper. 

We define a Verma module $M_\lambda$ with weight $\lambda$ as the representation of $\mathfrak{g}$ where no constraint is imposed beyond the relations of the Lie algebra. 
This means that the character $[M_\lambda]$ of the Verma module $M_\lambda$ is given by 
\begin{equation}
\label{chMso3}
    [M_\lambda] = \frac{x^{\lambda}}{1-x^{-2}} \, ,
\end{equation}
where the lowering operator has eigenvalue $-2$. In particular, a Verma module is always infinite-dimensional. It may happen that a Verma module contains other Verma modules. Here, this happens only when $\lambda \in \mathbb{Z}_{\geq 0}$, where the Verma module $M_{-\lambda -2}$ is included in the Verma module $M_{\lambda}$.
In that case, we can construct the quotient module $M_{\lambda}/M_{-\lambda -2}$, which is finite-dimensional and irreducible. We call this irreducible module $L_{\lambda}$, and its character is 
\begin{equation}
    [L_{\lambda}] =[M_{\lambda}] - [M_{-\lambda -2}] =  \frac{x^{\lambda} - x^{-\lambda-2}}{1-x^{-2}} =x^{\lambda} + x^{\lambda-2} + \cdots + x^{-\lambda} \, .  
\end{equation}
The dimension of the module is $\lambda+1$.\footnote{For $\lambda$ a positive integer, we can think of $\lambda$ as twice the spin.}
When $\lambda$ is not a positive integer, the Verma module $M_{\lambda}$ is irreducible, and therefore we set $L_{\lambda}=M_{\lambda}$. 

As we will see later, it is natural to introduce the Weyl vector $\rho = 1$ of the $\mathfrak{so}(3)$ Lie algebra, and the Weyl group $W = \{1,-1\}$. We further introduce the dot action of the Weyl group on the weight space through the formula
\begin{equation}
    w \cdot \lambda = w(\lambda + \rho) - \rho \, . 
\end{equation}
In our present simple set-up, we find $w \cdot \lambda = \lambda$ for $w=1$ and $w \cdot \lambda = - \lambda - 2$ for $w=-1$, so the character of the general irreducible module can be rewritten 
\begin{eqnarray}
    \left[L_{\lambda}\right] &=& \left[M_{(1) \cdot \lambda}\right] - \left[M_{(-1) \cdot \lambda}\right] \qquad  \mbox{for } \lambda \in \mathbb{N}  \, ,  \\ 
    \left[L_{\lambda}\right] &=& \left[M_{(1) \cdot \lambda}\right] \qquad \qquad \qquad \quad  \, \, \mbox{for } \lambda  \notin
    \mathbb{N} \, .  
\end{eqnarray}
We make several observations. Firstly, the representation theory of highest weight modules is a generalization of the representation theory of finite dimensional modules.
Secondly, integer weights behave differently from non-integer weights. More precisely, dominant weights give rise to the familiar Weyl character formula for finite dimensional representations, which involves a sum over the Weyl group of the algebra. Thirdly, we observe that the expression of the character of the irreducible module depends on the relative position of the weight $\lambda$ with respect to (minus) the Weyl vector $-\rho=-1$. All these observations  generalize to other semisimple Lie algebras.\footnote{For the exploration of other objects in the  category of highest weight modules of $\mathfrak{so}(3)$ in a physical context, see \cite{Troost:2012ck}.}

\subsection{\texorpdfstring{The Representations of the $\mathfrak{so}(2,3)$ Algebra }{}}
We  perform a similar analysis for the highest weight representations of the
$\mathfrak{so}(2,3)=B_2$ algebra. The choice of real form of the algebra does not matter at this stage,
but we must come back to this point when we consider the question of unitarity. We choose
a Cartan subalgebra in the compact subalgebra $\mathfrak{so}(2) \oplus \mathfrak{so}(3)$ of $\mathfrak{so}(2,3)$,
corresponding to the dilatation operator and a spin component. The Verma module $M_\lambda$ with highest-weight $\lambda$ generically has character 
\begin{equation}
\label{characterMlambdaB2}
    [M_{\lambda}] = \frac{z^{\lambda}}{\prod\limits_{\alpha > 0} (1-z^{-\alpha})} 
    \, , 
\end{equation}
generalizing (\ref{chMso3}), where the product over negative roots makes sure that we take into account the free action of the lowering operators on the highest weight state.
Depending on the weight $\lambda$, the module $M_{\lambda}$ may be reducible, and the character of the irreducible module $L_{\lambda}$ will differ from the Verma character (\ref{characterMlambdaB2}). This can happen only if another Verma module $M_{\mu}$ is a strict submodule of $M_{\lambda}$ for some weight $\mu$.

\subsubsection*{Integral regular weights}

\begin{figure}[t]
    \centering
    \includegraphics[width=.9\textwidth]{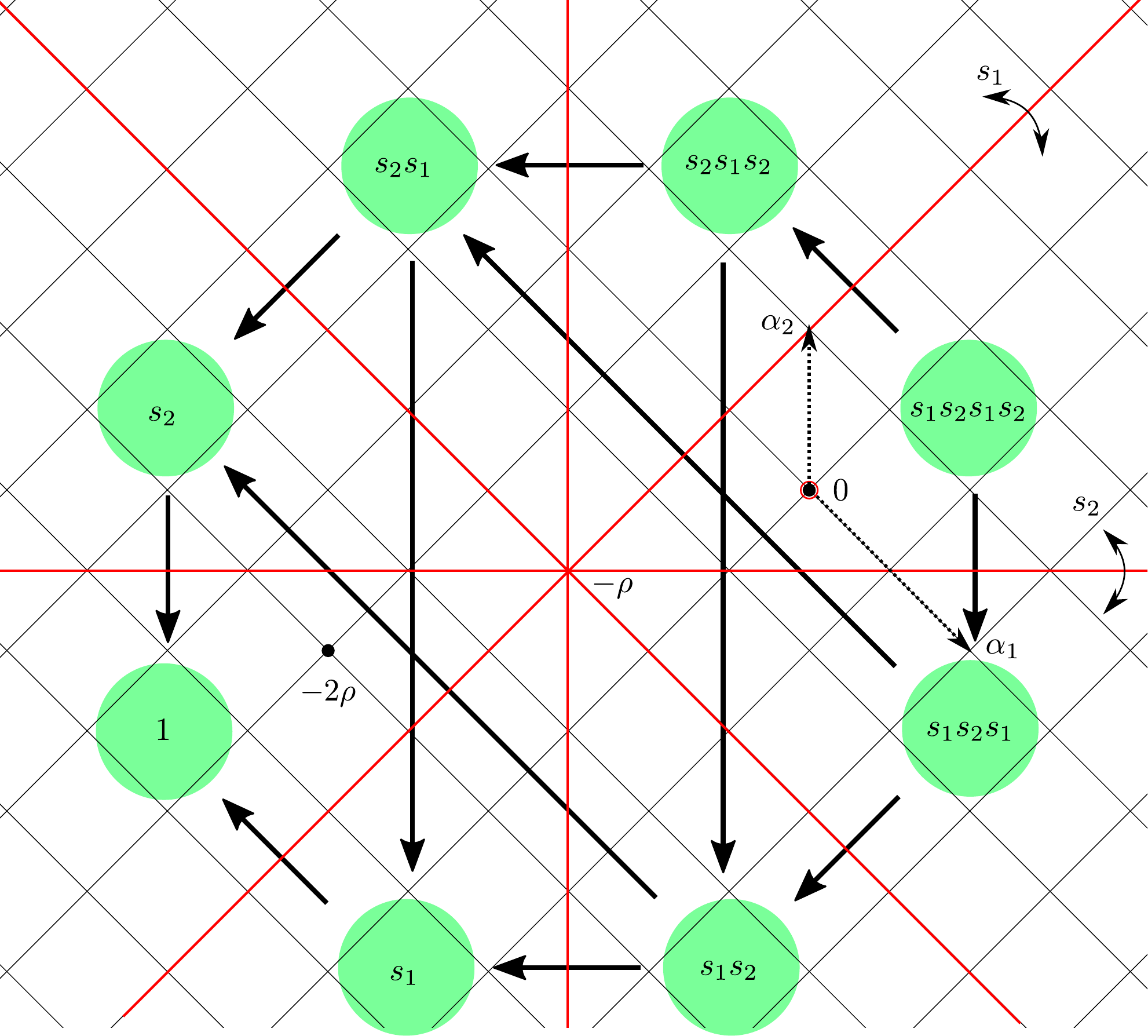}
    \caption{The $B_2$ (shifted) Weyl chambers, their associated Weyl group element in terms of simple reflections $s_i$, their Bruhat order, the simple roots $\alpha_i$ and the integral weight lattice. The red lines correspond to singular weights, and delimitate the shifted Weyl chambers. The intersections of gray lines correspond to integral weights. }
    \label{B2chambers}
\end{figure}
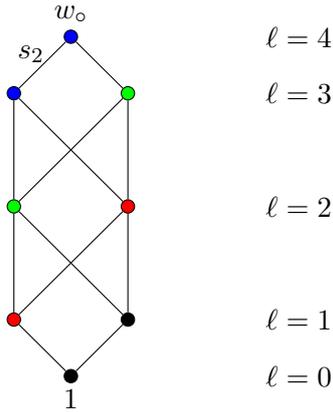
\begin{figure}[t]
    \centering
    \def\x{1.5}
    \def\y{1.5}
 \begin{tikzpicture}
\node[circle,fill=black,draw=black,inner sep=0pt,minimum size=5pt] (1) at (0.*\x,.5*\y) {};
\node[circle,fill=red,draw=black,inner sep=0pt,minimum size=5pt] (2) at (-0.5*\x,1*\y) {};
\node[circle,fill=black,draw=black,inner sep=0pt,minimum size=5pt] (3) at (0.5*\x,1*\y) {};
\node[circle,fill=green,draw=black,inner sep=0pt,minimum size=5pt] (4) at (-0.5*\x,2*\y) {};
\node[circle,fill=red,draw=black,inner sep=0pt,minimum size=5pt] (5) at (0.5*\x,2*\y) {};
\node[circle,fill=blue,draw=black,inner sep=0pt,minimum size=5pt] (6) at (-0.5*\x,3*\y) {};
\node[circle,fill=green,draw=black,inner sep=0pt,minimum size=5pt] (7) at (0.5*\x,3*\y) {};
\node[circle,fill=blue,draw=black,inner sep=0pt,minimum size=5pt] (8) at (0.*\x,3.5*\y) {};
\node[] (20) at (2.*\x,3.5*\y) {$\ell = 4$};
\node[] (20) at (2.*\x,3.0*\y) {$\ell = 3$};
\node[] (20) at (2.*\x,2.0*\y) {$\ell = 2$};
\node[] (20) at (2.*\x,1.0*\y) {$\ell = 1$};
\node[] (20) at (2.*\x,0.5*\y) {$\ell = 0$};
\node[] (9) at (-.35*\x,3.35*\y) {$s_2$};
\node[] (10) at (0*\x,3.7*\y) {$w_{\circ}$};
\node[] (11) at (0*\x,.3*\y) {$1$};
\draw[-] (2)-- (1);
\draw[-] (3)-- (1);
\draw[-] (4)-- (2);
\draw[-] (4)-- (3);\draw[-] (5)-- (2);\draw[-] (5)-- (3);\draw[-] (6)-- (4);\draw[-] (6)-- (5);\draw[-] (7)-- (4);\draw[-] (7)-- (5);\draw[-] (8)-- (6);\draw[-] (8)-- (7);
\end{tikzpicture}   
    \caption{The Bruhat order for the Weyl group of $B_2$, and the length function. The longest element $w_{\circ}$ has length four. There are two elements of length three. The lines at $45^{\circ}$ correspond to the $s_2$ reflection, and we use the same color for elements of $W$ connected by this reflection: they contribute to the same module $M^c$, see equation (\ref{compactVermaso5}). 
    }
    \label{B2Bruhat}
\end{figure}

Firstly, we consider for simplicity an integral weight $\lambda$: for each root $\alpha$ the product 
$\langle \lambda, \alpha^\vee \rangle$ satisfies $ \langle \lambda , \alpha^{\vee} \rangle \in \mathbb{Z}$. The weight space is two-dimensional, and the position of $\lambda$ with respect to the negative Weyl vector $- \rho$ is characterized by the sign of the integers $ \langle \lambda + \rho , \alpha^{\vee} \rangle$. This defines eight (shifted) Weyl chambers, as shown in Figure \ref{B2chambers}. One can label the chambers with elements of the Weyl group $W$, associating the identity element to the chamber that contains the weight $- 2 \rho$, as  in Figure \ref{B2chambers}.
On the Weyl group, one can define a partial order,  the Bruhat order \cite{HumphreysCoxeter}. This order can be summarized in a Bruhat graph represented in Figure \ref{B2Bruhat}, as well as on Figure \ref{B2chambers}. In a minimal representation of a Weyl group element in terms of simple reflections,
the length of the element is equal to the number of simple reflections.
Any integral weight in the interior of one of the (shifted) Weyl chambers can be written in a unique way as $\lambda = w \cdot \bl$, where $w \in W$ and $\bl$ is antidominant, meaning that $ \langle \bl + \rho , \alpha^{\vee} \rangle \notin \mathbb{Z}_{>0}$ for each positive root $\alpha$. 
 
 The partial Bruhat order is instrumental in our understanding of the structure of Verma modules \cite{humphreys2008representations}. Indeed, for an integral weight lying in the \emph{interior} of the antidominant Weyl chamber, and any Weyl group element $w$, we have that the irreducible module (and character) can  be understood in terms of the Verma modules (and characters) associated to the same antidominant weight, and Weyl group elements smaller than $w$ in the Bruhat order:
 \begin{equation}
  \label{KLB2}
     [L_{w \cdot \bl}] = \sum\limits_{w' \leq w} b_{w',w} [M_{w' \cdot \bl}]
 \end{equation}
 for some  integer coefficients $b_{w',w}$. 
 In the case of the algebra $\mathfrak{so}(2,3)$, these coefficients are particularly simple -- and it is mostly here  that we exploit that we restrict to the example of $\mathfrak{so}(3,2)$ in our warm-up section. The coefficients $b_{w',w}$ for the $\mathfrak{so}(2,3)$ algebra
 are given by 
 \begin{equation}
     b_{w',w} = (-1)^{\ell (w) - \ell (w')} \, , 
 \end{equation}
 where $\ell (w)$ is the length of the Weyl group element $w$, which can be read from the Bruhat graph \cite{HumphreysCoxeter} (see figures \ref{B2chambers} and \ref{B2Bruhat}). The dotted Weyl group action is still given by  the formula $w \cdot \lambda = w(\lambda + \rho) - \rho$. We have restricted to integral weights in the interior of a Weyl chamber -- those are called regular. We turn to an example.

 \paragraph{Example:} 
 Firstly, let us introduce the parameterization of roots and weights in terms of an orthonormal basis $e_i$ (described in detail in appendix \ref{conventions}).
 The simple roots are $\alpha_1=e_1-e_2$ and $\alpha_2=e_2$, see Figure \ref{B2chambers}. The $\mathfrak{so}(3,2)$ weights are denoted $(\lambda_1,\lambda_2)$ for a weight $\lambda=\lambda_1 e_1 + \lambda_2 e_2$.
 Let us then consider the weight $\lambda = (-1,2)$.
 It sits inside the Weyl chamber labeled by the Weyl group element $w=s_2 s_1 s_2$ of length three. For this example, the formula (\ref{KLB2}) gives rise to the
 character 
 \begin{equation}
     [L_{(-1,2)}] = [M_{(-1,2)}] - [M_{(-1,-3)}] - [M_{(-2,2)}]+ [M_{(-2,-3)}]+[M_{(-4,0)}]- [M_{(-4,-1)}] \, . 
 \end{equation}

 \subsubsection*{Integral singular weights}

The formula (\ref{KLB2}) provides the character of any irreducible highest-weight module with highest weight in the interior of a Weyl chamber, i.e. away from the red lines in Figure \ref{B2chambers}.  Now we  focus on singular integral weights, which are the integral weights $\lambda$ such that $ \langle \lambda + \rho , \alpha^{\vee} \rangle = 0$ for at least one root $\alpha$. They lie on a red line in Figure \ref{B2chambers}. The rule here is as follows : consider all the Weyl group elements that label the Weyl chambers  of which the closure contains $\lambda$, and pick the smallest such group element $w$ according to the Bruhat order. We can then write again $\lambda = w \cdot \bl$ for an antidominant weight $\bl$, and the character formula (\ref{KLB2}) remains true.

 \paragraph{Example:} 
 Let us consider the weight $\lambda = (- \frac{3}{2} , \frac{1}{2})$. This is an integral weight, but it is singular. It belongs to the closure of the Weyl chambers labeled by the Weyl group elements $s_2 s_1$ and $s_2 s_1 s_2$ of length two and three respectively. The smallest of these two elements is $w = s_2 s_1$, and therefore one  writes $\lambda = (- \frac{3}{2} , \frac{1}{2}) = w \cdot (- \frac{5}{2} , -\frac{1}{2})$. We then compute 
 \begin{equation}
     [L_{(- \frac{3}{2} , \frac{1}{2})}] = [M_{(- \frac{3}{2} , \frac{1}{2})}] - [M_{(- \frac{5}{2} , -\frac{1}{2})}] - [M_{(- \frac{3}{2} , -\frac{3}{2})}]+ [M_{(- \frac{5}{2} , -\frac{1}{2})}] =  [M_{(- \frac{3}{2} , \frac{1}{2})}] - [M_{(- \frac{3}{2} , -\frac{3}{2})}]  \, .
     \nonumber
 \end{equation}
 The cancellation between Verma module characters occurs  because we are studying a representation with singular highest weight.

 \subsubsection*{Non-integral weights}
 
Finally, we  extend our computation to non-integral weights. 
For an arbitrary weight $\lambda$, we construct  the set $\Phi_{[\lambda]}$ of roots $\alpha$ that satisfy $\langle \lambda , \alpha^{\vee} \rangle \in \mathbb{Z}$. To get a grasp on $\Phi_{[\lambda]}$, we compute this scalar product for all positive roots, with as before $\lambda = \lambda_1 e_1 + \lambda_2 e_2$. See Table \ref{tabintegralB2}. A priori, since there are four positive roots we have $2^4=16$ configurations to consider, but consistency restricts this number to 7 configurations, which are listed in Table \ref{tabintegralB2}. One observes that  $\Phi_{[\lambda]}$ is a root system, and its Weyl group $W_{[\lambda]}$ will play the role that the Weyl group $W$ of the whole algebra played in the integral case. The root system $\Phi_{[\lambda]}$ determines the \emph{integrality class} of $\lambda$. In this low rank case, the integer coefficients $b_{w,w'}$ again simplify to a sign depending on the length of the  elements in the group $W_{[\lambda]}$. The character formula takes  the form (\ref{KLB2}), but where the sum is restricted to the Weyl group elements $W_{[\lambda]}$ and the length function is inside this group. In this manner, we have found the characters of all irreducible highest weight representations of the $B_2$ algebra.

\begin{table}[]
    \centering
    \begin{tabular}{|c|c|c|c|c|c|c|c|c|}
    \hline
    $\alpha$ & $\langle \lambda , \alpha^{\vee} \rangle$ &  &    &  &   &   &   &  \\ \hline 
        $\alpha_1$ & $\lambda_1 - \lambda_2$ &  & $\mathbb{Z}$& & & & & $\mathbb{Z}$ \\
        $\alpha_2$ & $2 \lambda_2$ & & & $\mathbb{Z}$& & & $\mathbb{Z}$& $\mathbb{Z}$\\
        $\alpha_1 + \alpha_2$ & $2 \lambda_1$ & & & & $\mathbb{Z}$& & $\mathbb{Z}$& $\mathbb{Z}$\\
        $\alpha_1 + 2 \alpha_2$ & $\lambda_1 + \lambda_2$ & & & & & $\mathbb{Z}$ & & $\mathbb{Z}$ \\ \hline $\Phi_{[\lambda]}$
         &   & $1$ &  $A_1$ & $A_1$ & $A_1$ &$A_1$  & $D_2$ & $B_2$ \\ \hline 
    \end{tabular}
    \caption{The positive roots $\alpha$ of the $B_2$ algebra, the scalar product of the roots with
    the weights $\lambda=\lambda_1 e_1+\lambda_2 e_2$ as well as the root systems $\Phi_{[\lambda]}$ they give rise to.}
    \label{tabintegralB2}
\end{table}

\paragraph{Example: }
Consider the weight $\lambda = (- \frac{1}{2} , 0)$. The integrality class is $D_2$, and the associated Weyl group has four elements,  $W_{[\lambda]}=\{1 , s_2 , s_1s_2s_1 , s_1s_2s_1s_2\}$, using the notations of Figure \ref{B2chambers}. The weight $\lambda$ lies in the chamber of the longest element $s_1s_2s_1s_2$, so the irreducible character with highest weight $\lambda$ is
 \begin{equation}
     [L_{(- \frac{1}{2} ,  0)}] = [M_{(- \frac{1}{2} ,  0)}] - [M_{(- \frac{1}{2} ,  -1)}] - [M_{(- \frac{5}{2} ,  0)}]+ [M_{(- \frac{5}{2} ,  -1)}]  \, .
     \nonumber
 \end{equation}

\subsection{The Unitary Representations}
\label{subseqUniRepso3,2}

\begin{figure}[t]
    \centering
    \includegraphics[width=.9\textwidth]{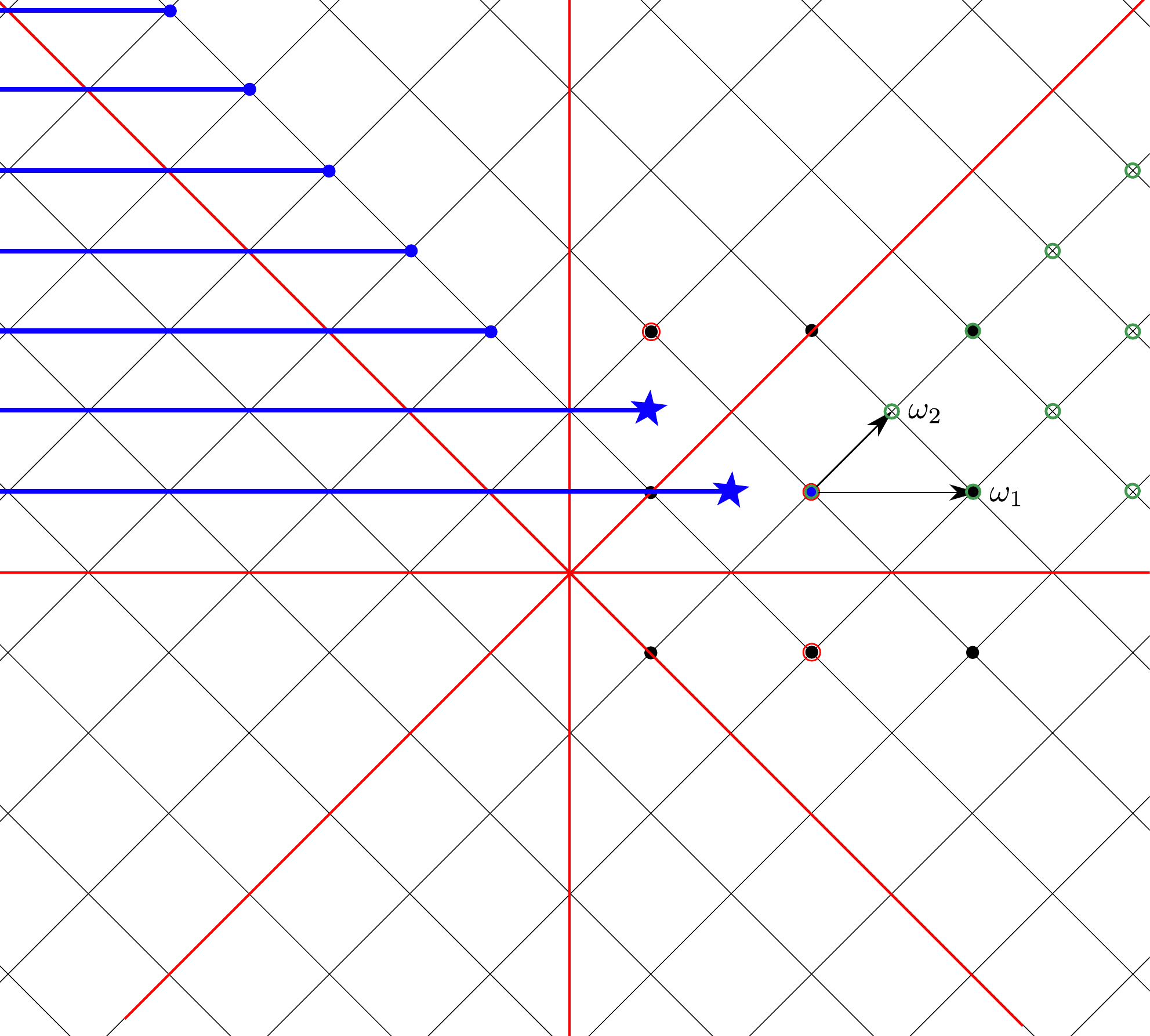}
    \caption{Weights in blue correspond to unitary representations of the three-dimensional 
    conformal algebra $\mathfrak{so}(2,3)$. The green circles correspond to dominant weights. }
    \label{B2unitarity}
\end{figure}

As we  review in full generality in section \ref{unitary}, only a subset of the irreducible
modules $L_{\lambda}$ are unitarizable. We  say that a weight is \emph{unitary} if the corresponding irreducible module $L_{\lambda}$ is unitarizable. In this context, it is important that we consider the real form $\mathfrak{so}(2,3)$ of the complex $B_2$ algebra. Manifestly, this is a non-compact real form, and therefore  non-trivial unitary representations will be infinite-dimensional. As we recall in section \ref{unitary}, in the case of the algebra $\mathfrak{so}(2,3)$, the result of the identification of unitary weights is as represented in Figure \ref{B2unitarity}, where the unitary weights are painted in blue. 

A second observation is that for all unitary weights we have that $2 \lambda_2 \in \mathbb{N}$. 
Thus, from the point of view of Table \ref{tabintegralB2}, the unitary weights correspond to the third, sixth or seventh cases, i.e. with root systems $\Phi_{[\lambda]} = A_1 , D_2$ or $B_2$. This corresponds to the fact that the compact subalgebra $\mathfrak{su}(2) = \mathfrak{so}(3) \subset \mathfrak{so}(2,3)$ is finitely represented in a unitary highest weight representation. In other words, for unitary irreducible modules the only source of infinite-dimensionality is the non-compact part of the algebra. We  exploit this fact to write more compact formulas for the characters. 
Firstly, we introduce  notations that reflect this desire.

For a unitary weight, let us define a module $M^c_{\lambda}$ which is the quotient of two Verma modules:
\begin{equation}
\label{compactVermaso5}
    M^c_{\lambda} = M_{\lambda} / M_{s_2 \cdot \lambda} \, . 
\end{equation}
This is sensible because of the restriction on unitary weights.
Accordingly, the character of the module $M^c_\lambda$ is 
\begin{equation}
    [M^c_{\lambda}] = [M_{\lambda}] -[M_{s_2 \cdot \lambda}] \, . 
\end{equation}
Thus, we have already divided out a Verma module that is guaranteed to be a submodule because of the fact that the compact algebra is finitely represented.
Using this notation, we can write down the characters of all irreducible unitary representations of $\mathfrak{so}(2,3)$ as follows: 
\begin{enumerate}
    \item[0)] The highest 
    weight $\lambda = 0$ corresponds to the trivial representation, and we simply have
    \begin{equation}
    [L_{\lambda}] = 1 \, .
\end{equation}
\item[1)] For highest weights $\lambda$ which fall in one of the following categories: 
\begin{itemize}
    \item $2 \lambda_1 \notin \mathbb{Z}$ (the $A_1$ case)
    \item $2 \lambda_1 \in \mathbb{Z}$ and $\lambda_1 - \lambda_2 \notin \mathbb{Z}$ (the $D_2$ case) and $\lambda_1 \leq - \frac{3}{2}$ (the weight is in the North-West chamber of $D_2$)
    \item $\lambda$ is integral (the $B_2$ case) and $\lambda_1 \leq - \lambda_2 - 2$ (the West-North chamber of $B_2$)
    \item $\lambda = \left( - \frac{3}{2} , \frac{1}{2} \right)$ or $\lambda = \left(-1,0\right)$ \, ,
\end{itemize}
we find that the compact subtraction is the end of the story
\begin{equation}
    [L_{\lambda}] =  [M^c_{\lambda}]     \, . 
\end{equation}
\item[2)] For $\lambda$ in one of the two following categories, we find a further subtraction: 
\begin{enumerate}
    \item[2a)] If $2 \lambda_1 \in \mathbb{Z}$ and $\lambda_1 - \lambda_2 \notin \mathbb{Z}$ (the $D_2$ case) and $\lambda_1 > - \frac{3}{2}$ (the weight is in the North-East chamber of $D_2$) -- this category contains only two weights, $\lambda = \left( - \frac{1}{2} , 0 \right)$ and $\lambda = \left( - 1  , \frac{1}{2} \right)$, 
\begin{equation}
    [L_{\lambda}] =  [M^c_{\lambda}] - [M^c_{(s_1 s_2 s_1) \cdot \lambda}]     \, . 
\end{equation}    
    \item[2b)] If $\lambda$ is integral (the $B_2$ case) and $\lambda_1 > -\lambda_2 - 2$ and $\lambda_1 < - \frac{3}{2}$ (the North-West chamber of $B_2$)
    \begin{equation}
    [L_{\lambda}] =  [M^c_{\lambda}] - [M^c_{(s_2 s_1 s_2) \cdot \lambda}]     \, . 
\end{equation}
\end{enumerate}
\end{enumerate}
These results comprise all characters of unitary irreducible highest weight representations of the 
conformal algebra $\mathfrak{so}(3,2)$. 
In the next subsection, we render more manifest the physical content of these results.

\subsection{In Physics Conventions}
Early physics references classifying the unitary representations of the
$\mathfrak{so}(3,2)$ algebra and their characters are \cite{Dirac:1963ta,Fronsdal:1965zzb,evans1967discrete} and \cite{dobrev1991spectrum}.
The algebra $\mathfrak{so}(2,3)$ admits a basis made of three $\mathfrak{so}(3)$ spins $J_{1,2,3}$, three translations $P_{1,2,3}$, three special conformal transformations $K_{1,2,3}$ and the dilatation operator $D$. In order to define the Verma modules, we declare two operators to be in the Cartan subalgebra,
which we choose to be the spin component $J_3$ and the dilatation operator $D$ which are in a compact
subalgebra. We pick four raising operators ($J_+$ and $K_{1,2,3}$) and four lowering operators ($J_-$ and $P_{1,2,3}$). We consider highest-weight modules, so the weights $\lambda$ will consist of eigenvalues $(-E, j)$ of $(-D , J_3)$. 
In these conventions, closer to traditions in physics, the above generic Verma module characters translate
into
\begin{equation}
\label{characterMlambdaB2physics}
    [M_{\lambda}] = \frac{z^{\lambda}}{\prod\limits_{\alpha > 0} (1-z^{-\alpha})} = \frac{x^{E} s^{2j}}{(1-s^{-2}) (1-x s^2)(1-x) (1-x s^{-2})  }  \, , 
\end{equation}
where the fugacity $x$ keeps track of the conformal dimension of the states, while the fugacity $s$ codes
(twice) the $3$-component of the spin.
The characters with respect to the $\mathfrak{su}(2)$ compact subalgebra read
\begin{equation}
\label{Mcso5}
    [M^c_{\lambda}] = \frac{x^{E} [L^{\mathfrak{su}(2)}_{2j}] }{(1-x s^2)(1-x) (1-x s^{-2})} \, . 
\end{equation}
with the usual spin $j$ character $[L^{\mathfrak{su}(2)}_{2j}]$
of the representation of the ${\mathfrak{su}(2)}$ subalgebra defined by
\begin{equation}
\label{su2finite}
    [L^{\mathfrak{su}(2)}_{2j}] = \sum\limits_{k=0,1,\dots}^{2j} s^{2(j-k)} \, . 
\end{equation}
On the lower blue line in figure \ref{B2unitarity}, we find the trivial representation with 
ground state energy and spin $(-E=\lambda_1,j=\lambda_2)=(0,0)$, the singleton $(-E,j)=(-1/2,0)$ 
as well as the other scalars  $(-E<-1/2,0)$.  On the second line, we have the singleton $(-1,1/2)$, as well as the other spinors $(-E<-1,1/2)$. The other representations are the generic $(-E \le -j-1,j)$ representations. See e.g. \cite{dobrev1991spectrum} for an early summary.

For the weights of type 1) in subsection \ref{subseqUniRepso3,2}, which include the generic scalar, spinor and higher spin representations we find the characters
\begin{equation}
    [L_{\lambda}] =  [M^c_{\lambda}] = \frac{x^{E} [L^{\mathfrak{su}(2)}_{2j}] }{(1-x s^2)(1-x) (1-x s^{-2})}\, .
\end{equation}
For the weights of type 2), we have for the singletons (type 2a)) 
\begin{equation}
    [L_{\lambda}] =  [M^c_{\lambda}] - [M^c_{(s_1 s_2 s_1) \cdot \lambda}] = \frac{x^{E} [L^{\mathfrak{su}(2)}_{2j}]  - x^{3-E} [L^{\mathfrak{su}(2)}_{2j}]  }{(1-x s^2)(1-x) (1-x s^{-2})} \, ,  
\end{equation}
and for the other extremal representations (type 2b) -- note that for those, $j \geq 1$), 
\begin{equation}
    [L_{\lambda}] =  [M^c_{\lambda}] - [M^c_{(s_2 s_1 s_2) \cdot \lambda}] = \frac{x^{E} [L^{\mathfrak{su}(2)}_{2j}]  - x^{E +1} [L^{\mathfrak{su}(2)}_{2j-2}]  }{(1-x s^2)(1-x) (1-x s^{-2})} \, . 
\end{equation}
These calculations exhaust the characters of unitary highest weight representations of $\mathfrak{so}(3,2)$, and are in agreement with the physics literature.

\subsection*{Summary Remarks}
The warm-up example of the three-dimensional conformal algebra is illuminating in multiple respects. It identifies the crucial role of the Weyl vector and the Weyl group for all irreducible characters, as well as the role of the compact subalgebra in the simplification of the unitary characters. It also motivated that we need to come to terms with at least two more advanced mathematical concepts: the first is the multiplicity of the Verma modules in the characters of irreducible modules, and the second is the generic classification of unitary highest weight representations. The generic treatment of these points requires  further levels of abstraction.

\section{The Characters of Irreducible Representations}
\label{characters}
In this section, we explain how to write the characters of irreducible modules in terms of the characters of Verma modules for an arbitrary semisimple complex Lie algebra $\mathfrak{g}$. 
Since the full mathematical solution to this problem is available, but may be hard to read, or even identify, we provide a very brief guide to the history and literature.

Important early contributions to the understanding of the category of modules with highest weight are \cite{bernstein1971structure}  and \cite{jantzen1979moduln}. The generic solution to the character
calculation is based on the Kazhdan-Luzstig conjecture \cite{kazhdan1979representations}
proven in \cite{BB,brylinski1981kazhdan}. The book \cite{humphreys2008representations} makes 
 the mathematics significantly more accessible. Furthermore, to understand the unitary characters the parabolic Kazhdan-Lusztig polynomials  \cite{Deodhar} are instrumental, in particular as pertaining
to Hermitian symmetric spaces \cite{ECartan}. The parabolic polynomials were computed in \cite{boe1988kazhdan} and in more technical detail in \cite{brenti2009parabolic}. The final step in summarizing the literature requires the use of translation functors \cite{humphreys2008representations}, and the resulting final formulation is most easily read in \cite{kashiwara1999characters}  and \cite{jantzen2008character}. We  refer to the book \cite{humphreys2008representations}  as well as to the  summary \cite{jantzen2008character} for  further history.

\subsection{The Kazhdan-Lusztig Theory}
\label{sectionKLTheory}
In this subsection, we briefly remind the reader of basic concepts in Lie algebra and representation
theory. See e.g. \cite{bourbaki1972groupes,humphreys2012introduction,OV} for gentler introductions.
Let $\mathfrak{g}$ be a semisimple complex Lie algebra, with Cartan subalgebra $\mathfrak{h}$. We denote the set of roots of $\mathfrak{g}$ by $\Phi$,  the subset of positive roots by $\Phi^+$ and by
$\Phi^s$ is the subset of simple roots. 
The Weyl group is $W$, the Weyl vector $\rho$, and we define the dot action 
\begin{equation}
    w \cdot \lambda = w (\lambda + \rho) - \rho \, . 
\end{equation}
Given a weight $\lambda \in \mathfrak{h}^\ast$, we define the root system $\Phi_{[\lambda]} = \{ \alpha \in \Phi | \langle \lambda , \alpha^{\vee} \rangle \in \mathbb{Z} \}$. Its Weyl group is denoted $W_{[\lambda]}$. The Bruhat order on $W_{[\lambda]}$ is consistent with the Bruhat order on W, and the parity of the length functions agree
\cite{HumphreysCoxeter}.

We will  use a handy parameterization for the weights \cite{humphreys2008representations}. A weight is called\footnote{We warn the reader that some authors  use  different definitions for these concepts. }
\begin{itemize}
    \item \emph{antidominant} if for all $\alpha \in \Phi^+$, $\langle \lambda + \rho , \alpha^{\vee} \rangle \notin \mathbb{Z}^{>0}$;
    \item \emph{dominant} if for all $\alpha \in \Phi^+$, $\langle \lambda + \rho , \alpha^{\vee} \rangle \in \mathbb{Z}^{>0}$. 
\end{itemize}
Both of these subsets of weights are highly restrictive, and in particular, their union does not include
all weights. Note also that our definition of dominant makes all dominant weights integral. 
 For any weight $\lambda \in \mathfrak{h}^\ast$, there is a unique antidominant weight in the dot orbit $W_{[\lambda]} \cdot \lambda$. Therefore, any weight $\lambda$ can be written in a unique way as 
\begin{equation}
\label{definitionbl}
   \lambda = w \cdot \bl
\end{equation}
with $\bl$ antidominant and $w \in W_{[\lambda]}$ of \emph{minimal length}.
The minimal length requirement ensures that the decomposition (\ref{definitionbl}) is unique.

Given a weight $\lambda \in \mathfrak{h}^{\ast}$, we  focus our attention on two modules, which are both highest-weight modules with highest weight $\lambda$. The first one is the \emph{Verma module} $M_{\lambda}$. It is defined as the module generated from a highest weight state by the action of all lowering operators.\footnote{More precisely, the relevant object here is the universal enveloping algebra $\mathcal{U}(\mathfrak{g})$, which can be thought as $\mathfrak{g}$ with an associative product such that the Lie bracket is  given by the commutator. We start with the one-dimensional $(\mathfrak{h} \oplus \mathfrak{n}^+)$-module $\mathbb{C}_{\lambda}$ (where the raising operators $\mathfrak{n}^+$ give zero and the Cartan $\mathfrak{h}$ acts according to the linear form $\lambda$), and form the tensor product with $\mathcal{U}(\mathfrak{g})$, $M(\lambda) = \mathcal{U}(\mathfrak{g}) \otimes_{\mathcal{U}(\mathfrak{h} \oplus \mathfrak{n}^+)} \mathbb{C}_{\lambda}$. }
Its character $[M_{\lambda}]$ follows  from the definition, 
\begin{equation}
\label{definitionM}
    [M_{\lambda}] = \frac{z^{\lambda}}{\prod\limits_{\alpha \in \Phi^+} (1-z^{- \alpha})} \, . 
\end{equation}
We  introduce the \emph{simple module} $L_{\lambda}$ (also called the \emph{irreducible module}), which is the unique simple quotient of $M_{\lambda}$. Writing down the character of the module $L_{\lambda}$ is a central task in this paper. 

Given an antidominant weight $\bl \in \mathfrak{h}^\ast$, our goal is  to understand how the character $[L_{w \cdot \bl}]$ of the irreducible module $L_{w \cdot \bl}$ decomposes into characters of Verma modules $[M_{\mu}]$. Only weights $\mu$ of the form $\mu \in W_{[\bl]} \cdot \bl$ can contribute \cite{humphreys2008representations}, so we can write 
\begin{equation}
\label{KLgeneric}
    [L_{w \cdot \bl}] = \sum\limits_{w' \in W_{[\bl]}}  (-1)^{\ell (w,w')} P^{W_{[\bl]}}_{w',w} (1) [M_{w' \cdot \bl}] \, ,  
\end{equation}
where $P^{W_{[\bl]}}_{w',w} (1)$ are  coefficients, and we have factored out the sign contribution of
the length difference $\ell (w,w')=\ell (w)-\ell (w')$. The coefficients $P^{W_{[\bl]}}_{w',w} (1)$ are the Kazhdan-Lusztig polynomials
$P^{W_{[\bl]}}_{w',w} (q)$ associated to the Weyl group $W_{[\bl]}$ and two elements $w'$ and $w$ of the group $W_{[\bl]}$, evaluated at $q=1$. In the next subsection, 
we give an algorithm to compute these polynomials. Note that we have presented a crucial property of the theory of representations and characters, namely that the coefficients only depend on the relevant Weyl group \cite{humphreys2008representations}. This property was surmised early and proven late in the development of the theory. It implies that extensive manipulations of Lie algebra generators can be summarized in the more efficient combinatorics of the Weyl group only.

\subsection{The Kazhdan-Lusztig Polynomials}
\label{sectionKLpoly}
We review one algorithm to compute
the Kazhdan-Lusztig polynomials for Coxeter groups (which includes all Weyl groups that we encounter) \cite{HumphreysCoxeter}. 
Firstly, one computes the Bruhat partial order, that we denote by $\leq$. 
Secondly, we proceed as follows. Let $x,w$ be two elements of the Coxeter group $W$. We are ultimately interested in the Kazhdan-Lusztig polynomial $P_{x,w}(q)$, but the algorithm requires to compute as well an auxiliary integer denoted $\mu (x,w)$. 
\begin{itemize}
    \item If $x=w$, set $P_{x,w}(q)=1$ and $\mu (x,w) = 0$. 
    \item If $x \nleq w$, set $P_{x,w}(q)=0$ and $\mu (x,w) = 0$.
    \item If $x \leq w$ and $x \neq w$, then let $s$ be a simple reflection such that $\ell (sw) < \ell (w)$. Let $c=0$ if $x \leq sx$, and $c=1$ otherwise. Then set (see the core of the existence proof provided in \cite{HumphreysCoxeter}, section 7.11)
    \begin{equation}
        P_{x,w}(q) = q^{1-c} P_{sx,sw}(q) + q^c P_{x,sw}(q) - \sum \mu (z,sw) q^{(\ell (w) - \ell (z))/2} P_{x,z}(q)
    \end{equation}
    where the sum runs over those $z \in W$ such that $z \leq sw$ and $sz \leq z$. Finally, define\footnote{In particular, note that if the degree of  the polynomial $P_{x,w} (q)$ is strictly less than $(\ell (w) - \ell(x) -1)/2$, then $ \mu (x,w) = 0$. }  
    \begin{equation}
        \mu (x,w) = \textrm{Coefficient of } q^{(\ell (w) - \ell(x) -1)/2} \textrm{ in } P_{x,w} (q) \, . 
    \end{equation}
\end{itemize}
Using the algorithm, we can compute all the Kazhdan-Lusztig polynomials for the Weyl groups $W$ appearing in the character formula (\ref{KLgeneric}). Thus, the proof \cite{BB,brylinski1981kazhdan} of the Kazhdan-Lusztig conjecture \cite{kazhdan1979representations} solves the problem of determining all characters of highest weight representations of semisimple Lie algebras.

\subsection{The Finite-dimensional Representations}
\label{sectionFiniteDim}

The reader may find comfort in recovering the Weyl character formula for finite-dimensional irreducible representations as a particular case of the vast generalization (\ref{KLgeneric}). The irreducible representation $L_{\lambda}$ of the simple Lie algebra $\mathfrak{g}$ is finite-dimensional if and only if its highest weight $\lambda$ is dominant (see subsection \ref{sectionKLTheory}). 

Let the weight $\lambda$ be dominant. Then we can write the weight $\lambda$ in the form $\lambda = w_{\circ} \cdot \bl$ with the weight $\bl$ antidominant and $w_{\circ}$ the longest element of the Weyl group. For all elements $x$ in the Weyl
group $W$, the Kazhdan-Lusztig polynomial $P_{x,w_{\circ}}(q)$ trivializes to 
$P_{x,w_{\circ}}(q)=1$ \cite{humphreys2008representations}. Therefore, for finite dimensional representations, the generic character
formula (\ref{KLgeneric}) simplifies to 
\begin{equation}
    [L_{\lambda}] = \sum\limits_{w' \in W} (-1)^{\ell (w_{\circ},w')} [M_{w' \cdot \bl}] \, ,
\end{equation}
which includes a sum over the whole Weyl group. Intuitively, the further the highest weight is from antidominance, the bigger the character sum. For finite representations, the sum has the maximal number of terms.

\subsubsection*{A Remark on Some Singular Integral Weights}
According to our definition, a dominant weight can not be singular. In fact, the integral weights located in the dominant shifted Weyl chamber (those that satisfy $\langle \lambda + \rho , \alpha^{\vee} \rangle \in \mathbb{Z}^{\geq 0}$ for all positive roots $\alpha$) are split into two categories: the dominant weights and the singular weights. An interesting consequence of the general formula (\ref{KLgeneric}) is that the character of an irreducible module $L(\lambda)$ where $\lambda$ belongs to the second category vanishes. This property is useful in simplifying character formulas.

\subsection{Examples}
The generic character formula captures (among others) the character of all highest weight representations of the conformal algebras $\mathfrak{so}(2,n)$. In the rest of the paper, we will mainly be interested in the unitary highest weight representations, which are a small subclass of all highest weight representations. These are the representations most evidently relevant in physical theories. Nevertheless, non-unitary representations can play a role in unitary theories with gauge symmetries, or in non-unitary theories of relevance to physics. Therefore, we want to make the point that the mathematical formalism that we reviewed also readily computes the characters of this much more general set of representations. To stress that point, we compute an example character which involves a slightly more complicated Kazhdan-Lusztig polynomial.

\subsubsection*{A $B_3$ Example}
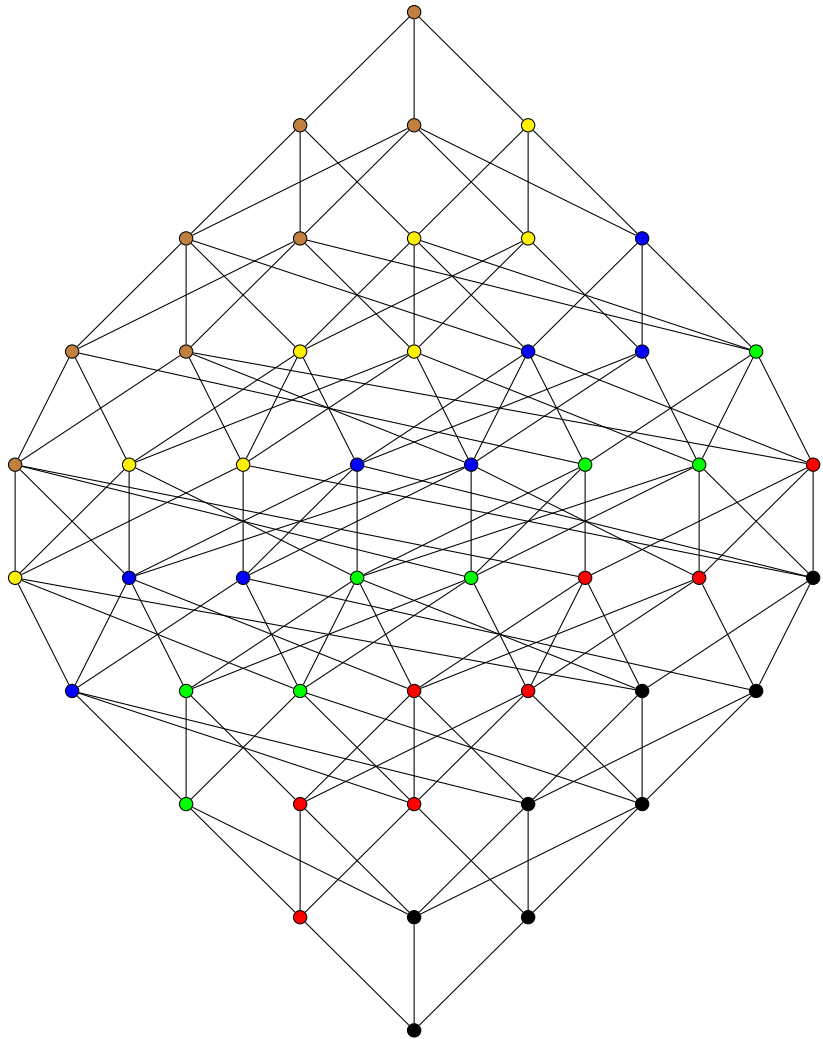
\begin{figure}[t]
    \centering
    \def\x{1.5}
    \def\y{1.5}
 \begin{tikzpicture}
\node[circle,fill=black,draw=black,inner sep=0pt,minimum size=5pt] (1) at (0.*\x,0*\y) {};\node[circle,fill=red,draw=black,inner sep=0pt,minimum size=5pt] (2) at (-1.*\x,1*\y) {};\node[circle,fill=black,draw=black,inner sep=0pt,minimum size=5pt] (3) at (0.*\x,1*\y) {};\node[circle,fill=black,draw=black,inner sep=0pt,minimum size=5pt] (4) at (1.*\x,1*\y) {};\node[circle,fill=green,draw=black,inner sep=0pt,minimum size=5pt] (5) at (-2.*\x,2*\y) {};\node[circle,fill=red,draw=black,inner sep=0pt,minimum size=5pt] (6) at (-1.*\x,2*\y) {};\node[circle,fill=red,draw=black,inner sep=0pt,minimum size=5pt] (7) at (0.*\x,2*\y) {};\node[circle,fill=black,draw=black,inner sep=0pt,minimum size=5pt] (8) at (1.*\x,2*\y) {};\node[circle,fill=black,draw=black,inner sep=0pt,minimum size=5pt] (9) at (2.*\x,2*\y) {};\node[circle,fill=blue,draw=black,inner sep=0pt,minimum size=5pt] (10) at (-3.*\x,3*\y) {};\node[circle,fill=green,draw=black,inner sep=0pt,minimum size=5pt] (11) at (-2.*\x,3*\y) {};\node[circle,fill=green,draw=black,inner sep=0pt,minimum size=5pt] (12) at (-1.*\x,3*\y) {};\node[circle,fill=red,draw=black,inner sep=0pt,minimum size=5pt] (13) at (0.*\x,3*\y) {};\node[circle,fill=red,draw=black,inner sep=0pt,minimum size=5pt] (14) at (1.*\x,3*\y) {};\node[circle,fill=black,draw=black,inner sep=0pt,minimum size=5pt] (15) at (2.*\x,3*\y) {};\node[circle,fill=black,draw=black,inner sep=0pt,minimum size=5pt] (16) at (3.*\x,3*\y) {};\node[circle,fill=yellow,draw=black,inner sep=0pt,minimum size=5pt] (17) at (-3.5*\x,4*\y) {};\node[circle,fill=blue,draw=black,inner sep=0pt,minimum size=5pt] (18) at (-2.5*\x,4*\y) {};\node[circle,fill=blue,draw=black,inner sep=0pt,minimum size=5pt] (19) at (-1.5*\x,4*\y) {};\node[circle,fill=green,draw=black,inner sep=0pt,minimum size=5pt] (20) at (-0.5*\x,4*\y) {};\node[circle,fill=green,draw=black,inner sep=0pt,minimum size=5pt] (21) at (0.5*\x,4*\y) {};\node[circle,fill=red,draw=black,inner sep=0pt,minimum size=5pt] (22) at (1.5*\x,4*\y) {};\node[circle,fill=red,draw=black,inner sep=0pt,minimum size=5pt] (23) at (2.5*\x,4*\y) {};\node[circle,fill=black,draw=black,inner sep=0pt,minimum size=5pt] (24) at (3.5*\x,4*\y) {};\node[circle,fill=brown,draw=black,inner sep=0pt,minimum size=5pt] (25) at (-3.5*\x,5*\y) {};\node[circle,fill=yellow,draw=black,inner sep=0pt,minimum size=5pt] (26) at (-2.5*\x,5*\y) {};\node[circle,fill=yellow,draw=black,inner sep=0pt,minimum size=5pt] (27) at (-1.5*\x,5*\y) {};\node[circle,fill=blue,draw=black,inner sep=0pt,minimum size=5pt] (28) at (-0.5*\x,5*\y) {};\node[circle,fill=blue,draw=black,inner sep=0pt,minimum size=5pt] (29) at (0.5*\x,5*\y) {};\node[circle,fill=green,draw=black,inner sep=0pt,minimum size=5pt] (30) at (1.5*\x,5*\y) {};\node[circle,fill=green,draw=black,inner sep=0pt,minimum size=5pt] (31) at (2.5*\x,5*\y) {};\node[circle,fill=red,draw=black,inner sep=0pt,minimum size=5pt] (32) at (3.5*\x,5*\y) {};\node[circle,fill=brown,draw=black,inner sep=0pt,minimum size=5pt] (33) at (-3.*\x,6*\y) {};\node[circle,fill=brown,draw=black,inner sep=0pt,minimum size=5pt] (34) at (-2.*\x,6*\y) {};\node[circle,fill=yellow,draw=black,inner sep=0pt,minimum size=5pt] (35) at (-1.*\x,6*\y) {};\node[circle,fill=yellow,draw=black,inner sep=0pt,minimum size=5pt] (36) at (0.*\x,6*\y) {};\node[circle,fill=blue,draw=black,inner sep=0pt,minimum size=5pt] (37) at (1.*\x,6*\y) {};\node[circle,fill=blue,draw=black,inner sep=0pt,minimum size=5pt] (38) at (2.*\x,6*\y) {};\node[circle,fill=green,draw=black,inner sep=0pt,minimum size=5pt] (39) at (3.*\x,6*\y) {};\node[circle,fill=brown,draw=black,inner sep=0pt,minimum size=5pt] (40) at (-2.*\x,7*\y) {};\node[circle,fill=brown,draw=black,inner sep=0pt,minimum size=5pt] (41) at (-1.*\x,7*\y) {};\node[circle,fill=yellow,draw=black,inner sep=0pt,minimum size=5pt] (42) at (0.*\x,7*\y) {};\node[circle,fill=yellow,draw=black,inner sep=0pt,minimum size=5pt] (43) at (1.*\x,7*\y) {};\node[circle,fill=blue,draw=black,inner sep=0pt,minimum size=5pt] (44) at (2.*\x,7*\y) {};\node[circle,fill=brown,draw=black,inner sep=0pt,minimum size=5pt] (45) at (-1.*\x,8*\y) {};\node[circle,fill=brown,draw=black,inner sep=0pt,minimum size=5pt] (46) at (0.*\x,8*\y) {};\node[circle,fill=yellow,draw=black,inner sep=0pt,minimum size=5pt] (47) at (1.*\x,8*\y) {};\node[circle,fill=brown,draw=black,inner sep=0pt,minimum size=5pt] (48) at (0.*\x,9*\y) {};\draw[-] (2)-- (1);\draw[-] (3)-- (1);\draw[-] (4)-- (1);\draw[-] (5)-- (2);\draw[-] (5)-- (3);\draw[-] (6)-- (2);\draw[-] (6)-- (3);\draw[-] (7)-- (2);\draw[-] (7)-- (4);\draw[-] (8)-- (3);\draw[-] (8)-- (4);\draw[-] (9)-- (3);\draw[-] (9)-- (4);\draw[-] (10)-- (5);\draw[-] (10)-- (7);\draw[-] (10)-- (8);\draw[-] (11)-- (5);\draw[-] (11)-- (6);\draw[-] (12)-- (5);\draw[-] (12)-- (7);\draw[-] (12)-- (9);\draw[-] (13)-- (6);\draw[-] (13)-- (7);\draw[-] (13)-- (8);\draw[-] (14)-- (6);\draw[-] (14)-- (7);\draw[-] (14)-- (9);\draw[-] (15)-- (8);\draw[-] (15)-- (9);\draw[-] (16)-- (8);\draw[-] (16)-- (9);\draw[-] (17)-- (10);\draw[-] (17)-- (12);\draw[-] (17)-- (15);\draw[-] (18)-- (10);\draw[-] (18)-- (11);\draw[-] (18)-- (13);\draw[-] (19)-- (10);\draw[-] (19)-- (12);\draw[-] (19)-- (16);\draw[-] (20)-- (11);\draw[-] (20)-- (12);\draw[-] (20)-- (13);\draw[-] (20)-- (15);\draw[-] (21)-- (11);\draw[-] (21)-- (12);\draw[-] (21)-- (14);\draw[-] (22)-- (13);\draw[-] (22)-- (14);\draw[-] (22)-- (15);\draw[-] (23)-- (13);\draw[-] (23)-- (14);\draw[-] (23)-- (16);\draw[-] (24)-- (15);\draw[-] (24)-- (16);\draw[-] (25)-- (17);\draw[-] (25)-- (18);\draw[-] (25)-- (21);\draw[-] (25)-- (22);\draw[-] (26)-- (17);\draw[-] (26)-- (18);\draw[-] (26)-- (20);\draw[-] (27)-- (17);\draw[-] (27)-- (19);\draw[-] (27)-- (24);\draw[-] (28)-- (18);\draw[-] (28)-- (19);\draw[-] (28)-- (20);\draw[-] (28)-- (24);\draw[-] (29)-- (18);\draw[-] (29)-- (19);\draw[-] (29)-- (21);\draw[-] (29)-- (23);\draw[-] (30)-- (20);\draw[-] (30)-- (21);\draw[-] (30)-- (22);\draw[-] (31)-- (20);\draw[-] (31)-- (21);\draw[-] (31)-- (23);\draw[-] (31)-- (24);\draw[-] (32)-- (22);\draw[-] (32)-- (23);\draw[-] (32)-- (24);\draw[-] (33)-- (25);\draw[-] (33)-- (26);\draw[-] (33)-- (30);\draw[-] (34)-- (25);\draw[-] (34)-- (27);\draw[-] (34)-- (29);\draw[-] (34)-- (32);\draw[-] (35)-- (26);\draw[-] (35)-- (27);\draw[-] (35)-- (28);\draw[-] (36)-- (26);\draw[-] (36)-- (27);\draw[-] (36)-- (29);\draw[-] (36)-- (31);\draw[-] (37)-- (28);\draw[-] (37)-- (29);\draw[-] (37)-- (30);\draw[-] (37)-- (32);\draw[-] (38)-- (28);\draw[-] (38)-- (29);\draw[-] (38)-- (31);\draw[-] (39)-- (30);\draw[-] (39)-- (31);\draw[-] (39)-- (32);\draw[-] (40)-- (33);\draw[-] (40)-- (34);\draw[-] (40)-- (35);\draw[-] (40)-- (37);\draw[-] (41)-- (33);\draw[-] (41)-- (34);\draw[-] (41)-- (36);\draw[-] (41)-- (39);\draw[-] (42)-- (35);\draw[-] (42)-- (36);\draw[-] (42)-- (37);\draw[-] (42)-- (39);\draw[-] (43)-- (35);\draw[-] (43)-- (36);\draw[-] (43)-- (38);\draw[-] (44)-- (37);\draw[-] (44)-- (38);\draw[-] (44)-- (39);\draw[-] (45)-- (40);\draw[-] (45)-- (41);\draw[-] (45)-- (42);\draw[-] (46)-- (40);\draw[-] (46)-- (41);\draw[-] (46)-- (43);\draw[-] (46)-- (44);\draw[-] (47)-- (42);\draw[-] (47)-- (43);\draw[-] (47)-- (44);\draw[-] (48)-- (45);\draw[-] (48)-- (46);\draw[-] (48)-- (47);
\end{tikzpicture}   
    \caption{The Bruhat order for $B_3$. As in Figure \ref{B2Bruhat}, we use the same color for elements of $W$ that contribute to the same module $M^c$ (see section \ref{factorization}). The compact subgroup $W_J$ is isomorphic to the Weyl group of $B_2$, and one can check that the subset of elements in each given color is isomorphic to Figure \ref{B2Bruhat}.  }
    \label{B3Bruhat}
\end{figure}

The Weyl group of $B_3$ has 48 elements. They are arranged in ten levels, depending on the number of 
simple Weyl reflections that occur in their reduced expression. 
See Figure \ref{B3Bruhat}. 
Since there are 48 Weyl chambers, and a proliferation of walls and weights of various singular types,
we do not provide a complete catalogue of characters. The results are straightforward to obtain, but unwieldy to present. We only provide a flavour of what such a catalogue looks like.

To discern the features of the catalogue, it is sufficient to 
analyze the geometry of the chambers, the walls, and the corners. 
 The positive root system
$\Phi^+=\{ e_1-e_2,e_2-e_3,e_3,e_2,e_1,e_1-e_3,e_1+e_2,e_1+e_3,e_2+e_3 \}$ of the algebra
$\mathfrak{so}(5,2)$ can be divided  into
subsystems in various ways. If the set of roots orthogonal to the weight $\lambda+\rho$
is empty, we are in a chamber. If it is non-empty, we are
on at least one wall. 
We have nine walls, given by the equations
$\lambda_i=\lambda_j$, $\lambda_i=-\lambda_j$ and $\lambda_i=0$. We have 
weights living on a single wall, 
weights living in the corner of two walls,
in the corner of three, 
in the corner of four 
or on the intersection of the nine walls. 
This provides us with a first glimpse of the structure of the catalogue.

Next, we want to clarify the difficulty of computing the Kazhdan-Lusztig polynomials. While
laborious, the difficulty remains well within reach of ancient computers.
The most complicated Kazhdan-Lusztig polynomial turns out to be $P_{1,s_2 s_3 s_2 s_1 s_2 s_3 s_2 }(q)=q^2+q+1$ (and it arises for a single other combination of Weyl group elements as well). 
At $q=1$, this will give rise to a triple multiplicity for a Verma module character in the character
sum formula. An example weight for which we need this polynomial is produced by acting with $w=
s_2 s_3 s_2 s_1 s_2 s_3 s_2$ 
on an anti-dominant weight. Thus, we give the following example entry in the catalogue.
\subsection*{Example}
Consider the weight $(s_2 s_3 s_2 s_1 s_2 s_3 s_2)\cdot (-2 \rho)=(-1,1,-1)$. We apply the general procedure outlined in this section using a symbolic manipulation program, and find the character:
\begin{eqnarray}
[L_{(-1,1-1)}] &=& 
-3 [M_{({-5,-3,-1})}]+2 [M_{({-5,-3,0})}] +3 [ M_{({-5,-2,-2})} ]
\nonumber \\ & & 
-2 [M_{({-5,-2,1})}]-2 [M_{({-5,-1,-2})}]+2 [M_{({-5,-1,1})}]
\nonumber \\ & & 
+2 [M_{({-5,0,-1})}]-2 [M_{({-5,0,0})}]+2 [M_{({-4,-4,-1})}]
\nonumber \\ & & 
-2 [M_{({-4,-4,0})}]-2 [M_{({-4,-2,-3})}]+[M_{({-4,-2,2})}]
\nonumber \\ & & 
+2 [M_{({-4,-1,-3})}]-[M_{({-4,-1,2})}]-[M_{({-4,1,-1})}]
\nonumber \\ & & 
+[M_{({-4,1,0})}]-2 [M_{({-3,-4,-2})}]+2 [M_{({-3,-4,1})}]
\nonumber \\ & & 
+2 [M_{({-3,-3,-3})}]-[M_{({-3,-3,2})}]-2 [M_{({-3,0,-3})}]
\nonumber \\ & & 
+[M_{({-3,0,2})}]+[M_{({-3,1,-2})}]-[M_{({-3,1,1})}]
\nonumber \\ & & 
+[M_{({-2,-4,-2})}]-[M_{({-2,-4,1})}]-[M_{({-2,-3,-3})}]
\nonumber \\ & & 
+[M_{({-2,-3,2})}]+[M_{({-2,0,-3})}]-[M_{({-2,1,-2})}]
\nonumber \\ & & 
-[M_{({-1,-4,-1})}]+[M_{({-1,-4,0})}]+[M_{({-1,-2,-3})}]
\nonumber \\ & & 
-[M_{({-1,-2,2})}]-[M_{({-1,-1,-3})}]+[M_{({-1,1,-1})}] \, .
\end{eqnarray}
Note the multiplicities of the Verma modules, which go up to three, even in this low rank example.
Proceeding in this fashion, one can imagine filling out systematically the  thick catalogue of character
formulas. The reader who is so inclined will surely find the tables to be constructed shortly equally mesmerizing.

\section{The Unitary Conformal Multiplets}
\label{unitary}
In section \ref{characters} we exhibited how to compute the structure and character of any highest weight representation
of the conformal algebra $\mathfrak{so}(2,n)$. 
In this section, we determine which of the highest weight conformal multiplets are unitary.
Those multiplets are the representation theoretic building blocks of unitary conformal field theories. The mathematical
analysis of the unitarizability of the representations of the conformal algebra fits into a more general framework,
which we recall  briefly.

Firstly, let $G$ be a simply connected and connected simple Lie group, and $K$ a closed maximal subgroup. Then, the group $G$ admits a non-trivial unitary highest weight module precisely when  $(G,K)$ is a Hermitian symmetric pair \cite{HC1951,HC1956}. 
Hermitian symmetric pairs have been classified \cite{ECartan}. See appendix \ref{realsimple} for
a summary of the relevant structure theory of real simple Lie groups, 
and \cite{knapp2013lie} for a complete treatment.
 The conformal group $G=\mathrm{SO}(2,n)$ satisfies the condition, with the maximal compact subgroup 
 $K=\mathrm{SO}(2) \times \mathrm{Spin}(n)$. The techniques used to classify the unitary highest weight representations
 for such groups include the identification of weights of null vectors and the degeneration of the contravariant form on the Verma module \cite{enright1983classification,jakobsen1983hermitian}.

The full classification of the unitary highest weight modules of the conformal algebras was obtained in \cite{enright1983classification}. It is based on an exploitation of necessary and sufficient inequalities satisfied by unitary representations. These were derived in full generality in \cite{parthasarathy1980criteria}. The analysis of physicists of level one and level two constraints on unitary representations can be viewed as a partial analysis of the necessary conditions. In this section, we demonstrate that it suffices to decipher the earlier and more complete mathematical classification results to recuperate  in a uniform manner the results in the physics literature. We provide a glimpse of the concepts that underlie the classification result, illustrate the general analysis in the example of $B_2=\mathfrak{so}(2,3)$, and then recall the full classification of the unitary highest weight multiplets for the $B_k=\mathfrak{so}(2,2k-1)$ and $D_k=\mathfrak{so}(2,2k-2)$ algebras. A physics reference in the same vein is \cite{Ferrara:2000nu}.

\subsection{Useful Concepts}
\begin{figure}[t]
\begin{center}
\begin{tikzpicture}
\node (minus) at (-0.8,0) [circle] {};
\node (zero) at (0,0)  [circle] {};
\node (one) at  (4,0) [circle,draw,fill,scale=0.3] {};
\node (1) at  (4,.4)  {$A(\Lambda)$};
\node (two) at (5,0) [circle,draw,fill,scale=0.3] {};
\node (three) at (6,0) [circle,draw,fill,scale=0.3] {};
\node (four) at (7,0) [circle,draw,fill,scale=0.3] {};
\node (4) at  (7,.4)  {$B(\Lambda)$};
\draw (zero.east) -- (one.west);
\draw[dotted,line width=.5pt] (minus.east) -- (zero.west);
\draw [
    thick,
    decoration={
        brace,
        mirror,
        raise=0.1cm
    },
    decorate
] (two) -- (three) 
node [pos=0.5,anchor=north,yshift=-4] {$C(\Lambda)$}; 
\end{tikzpicture}
\end{center}
\caption{The positioning of unitary highest weight representations in highest weight space. The highest weights lie on lines of the form $\lambda = \Lambda + z \zeta$, and the figure represents the values of $z \in \mathbb{R}$ that give unitary weights. On a given line, there is a semi-infinite line of highest weights which is allowed, and then an equally spaced set of discrete allowed values, starting at the end of the half-line, and ending after a finite number of steps. }
\label{positioningunitaries}
\end{figure}
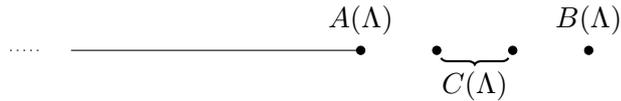

We again consider highest weight modules based on a highest weight state with respect to a Borel subalgebra $\mathfrak{b}$
of the complexified Lie algebra. The elements $h$ in the Cartan subalgebra $\mathfrak{h}$ act as $\lambda(h)$
where $\lambda$ is the highest weight. The span of the compact root system $\Phi_c$ has co-dimension one in the dual $\mathfrak{h}^\ast$ of the Cartan subalgebra \cite{knapp2013lie}. We define $\beta$ to be the maximal non-compact root \cite{knapp2013lie}. 
The classification theorem of \cite{enright1983classification} introduces a class of weights, which we generically write $\Lambda$,
which are $\Phi_c^+$ dominant (because the compact subalgebra $\mathfrak{k}$ is finitely represented) and 
which satisfy 
\begin{equation}
\langle \Lambda + \rho , \beta \rangle = 0 \, ,
\end{equation}
where $\beta$ is the maximal non-compact positive root of the conformal algebra.\footnote{In \cite{enright1983classification}, the weights $\Lambda$ are called $\lambda_0$. } We also introduce an 
element $\zeta$ of the weight space which satisfies that it is orthogonal to all compact roots as well as the normalization 
\begin{equation}
\frac{2 \langle \zeta, \beta \rangle}{\langle \beta , \beta \rangle} = 1 \, .
\end{equation}
Then the highest weights corresponding to unitarizable representations all lie on the lines
$\lambda = \Lambda + z \zeta$ where $z$ is a real number. See figure \ref{positioningunitaries}.

 There is a half-line of unitary representations ending at a point  which is generically
 at a positive value of $z$, depending on the algebra $\mathfrak{g}$ and the weight $\Lambda$. Then, there are further points where unitary representations can occur, taking values in an equally spaced set, with a spacing  which depends on the algebra only. There is an endpoint to this discrete set. The calculation of the three constants (called $A(\Lambda)$, $B(\Lambda)$ and $C(\Lambda)$ in Figure \ref{positioningunitaries}) that determine this set proceeds via the introduction of auxiliary root systems.

Indeed, we want to bring to the fore
how singular the weight $\Lambda$ is with respect to the compact root system. To that end, we define the subset $\Phi_c(\Lambda)$ of compact roots orthogonal to $\Lambda$. We then define the new root system $\{ \pm \beta \} \cup \Phi_c(\Lambda)$ and decompose it into simple root subsystems. The simple root system which contains the maximal non-compact root $\beta$ is baptized $Q(\Lambda)$.  Exceptionally, we will make use of a second root system $R(\Lambda)$, defined as follows. If the root system $\Phi$ has two root lengths and there is a short root not orthogonal to the system $Q(\Lambda)$ and such that $\langle \Lambda,\alpha \rangle/\langle \alpha,\alpha \rangle=1$, then we adjoin the short root to $Q(\Lambda)$.  The simple component containing
$\beta$ of the resulting root system is named $R(\Lambda)$. These root systems can  be algorithmically determined from the weight $\Lambda$, and they allow for the calculations of the three constants, which in turn determine all the unitary highest weight conformal multiplets. The calculations are performed explicitly in \cite{enright1983classification}. We  review the results of the calculations in subsections \ref{unitaryodd} and \ref{unitaryeven}.

\subsection{The Algebras \texorpdfstring{$B_k=\mathfrak{so}(2,2k-1)$}{}}
\label{unitaryodd}
In this subsection, we generalize the example of the $\mathfrak{so}(2,3)$ algebra to include all conformal algebras $\mathfrak{so}(2,2k-1)$ associated to a space-times
of odd dimension $2k-1$. We list all unitary highest weight representations \cite{enright1983classification}. We again use the conventions of Appendix \ref{conventions}. 
Firstly, we review the set $\Phi_c^+$ of positive compact roots, as well as the set
$\Phi_n^+$ of positive non-compact roots \cite{knapp2013lie}:
\begin{eqnarray}
\Phi^+_c &=& \{ e_i \pm e_j | 2 \le i < j \le n \} \cup \{ e_j | 2 \le j \le n \}
\nonumber \\
\Phi^+_n &=& \{ e_1 \pm e_j | 2 \le j \le n \} \cup \{ e_1 \} \, .
\end{eqnarray}
The highest non-compact root
\begin{equation}
\beta = e_1+e_2 
\end{equation}
coincides with the highest root of the algebra. When we include the highest root of the algebra in the Dynkin diagram, we obtain the affine untwisted Dynkin diagram.
The Weyl vector $\rho$ for the $B_k$ algebra is
\begin{equation}
\rho=(k-1/2,k-3/2,\dots,1/2) \, .
\end{equation}
We parameterize the weights $\Lambda$ in terms of their components in the basis $e_i$ of orthonormal vectors
and demand that the components corresponding to the compact subalgebra $\mathfrak{so}(2k-1)$ are dominant integral weights:
\begin{eqnarray}
\Lambda = (\lambda_1,\dots,\lambda_k) & & 
\nonumber \\
\lambda_2 \ge \lambda_3 \ge \dots \ge \lambda_k \ge 0 & & 
\nonumber \\
\lambda_i-\lambda_j \in \mathbb{Z} & & \quad \mbox{for} \quad 2\le i<j \le k 
\nonumber \\
2 \lambda_k \in \mathbb{Z} \, . &&
\end{eqnarray}
We moreover need the $\lambda_1$ component of the weight $\Lambda$ to be tuned such that
the weight $\Lambda+\rho$ is orthogonal to the maximal non-compact root $\beta$, which implies
\begin{equation}
\lambda_1+\lambda_2 = -2k+2 \, .
\end{equation}
We moreover parameterize the line on which the unitaries with highest weight $\lambda$ lie in terms of the normalized
orthogonal vector $\zeta$
\begin{eqnarray}
\zeta &=& (1,0,\dots,0)
\nonumber \\
\lambda &=& \Lambda + z \zeta \, .
\end{eqnarray}

The root systems $Q(\Lambda)$ and $R(\Lambda)$ that measure the degenerate nature of the anker weight $\Lambda$  will either be the full conformal algebra $\mathfrak{so}(2,n)$ or $\mathfrak{su}(1,p)$ with $p$ smaller than the rank of the conformal algebra. We distinguish three cases for the root systems $Q$ and $R$ \cite{enright1983classification}. The first case is labelled by an extra integer $p$ that satisfies $1 \le p < k$.
Case (I,$p$) corresponds to root system $Q=\mathfrak{su}(1,p)=R$ with anker weights $\Lambda$ obeying 
$\lambda_2=\lambda_3=\dots=\lambda_{p+1}$ for $p \le k-1$. Case II corresponds to $Q=\mathfrak{so}(2,2k-1)=R$ and  $\lambda_2=\dots=0$. Case III is exceptional in that it has a root system $Q=\mathfrak{su}(1,k-1)$ that differs from the root system
$R=\mathfrak{so}(2,2k-1)$. The weight satisfies $\lambda_2=\dots =1/2$.

The theorem of \cite{enright1983classification} states that the highest weight irreducible module with highest weight $\lambda=\Lambda+z \zeta$
is unitarizable if the module is trivial, or if the highest weight obeys the inequalities
\begin{eqnarray}
z \le p  & &  \mbox{ for case (I,p) and}
\nonumber \\
z \le k-1/2  &&  \mbox{ for cases II and III} \, .
\end{eqnarray}
Preparing for a physicist's energetic lowest weight perspective, we denote the first component of the weight $\lambda$ by $\lambda_1 = -E$. We summarize the unitarity conditions for $\mathfrak{so}(2,2k-1)$ in Tables
\ref{tabUnitaryBk} and \ref{TableUnitarityBk}.

\afterpage{
\begin{table}[]
\def\x{1}
    \centering
    \begin{tabular}{|c|c|}
    \hline
       (I,1)  & 
\begin{tikzpicture}
\node[circle,fill=black,inner sep=0pt,minimum size=10pt,label=below:{$0$}] (0) at (-.707*\x,.707*\x) {};
\node[circle,draw=black,inner sep=0pt,minimum size=10pt,label=below:{$1$}] (1) at (-.707*\x,-.707*\x) {};
\node[circle,draw=black,inner sep=0pt,minimum size=10pt,label=below:{$2$}] (2) at (0*\x,0*\x) {};
\node[circle,draw=black,inner sep=0pt,minimum size=10pt,label=below:{$3$}] (3) at (1*\x,0*\x) {};
\node (4) at (2*\x,0*\x) {...};
\node[circle,draw=black,inner sep=0pt,minimum size=10pt,label=below:{$p$}] (5) at (3*\x,0*\x) {};
\node[circle,draw=black,inner sep=0pt,minimum size=10pt,label=below:{$p+1$}] (6) at (4*\x,0*\x) {};
\node (7) at (5*\x,0*\x) {...};
\node[circle,draw=black,inner sep=0pt,minimum size=10pt,label=below:{$k-1$}] (8) at (6*\x,0) {};
\node[circle,draw=black,inner sep=0pt,minimum size=10pt,label=below:{$k$}] (9) at (7*\x,0) {};
\draw[-] (0) -- (2);
\draw[-] (1) -- (2);
\draw[-] (2) -- (3);
\draw[-] (3) -- (4);
\draw[-] (5) -- (4);
\draw[-] (5) -- (6);
\draw[-] (6) -- (7);
\draw[-] (8) -- (7);
\draw[transform canvas={yshift=2*\x}] (9) -- (8);
\draw[transform canvas={yshift=-2*\x}] (9) -- (8);
\draw (-.707*\x,.707*\x) circle (.3*\x);
\draw[-] (6.6*\x,0*\x) -- (6.4*\x,.2*\x);
\draw[-] (6.6*\x,0*\x) -- (6.4*\x,-.2*\x);
\end{tikzpicture}\\ \hline
          \begin{tabular}{c}
           (I,$p$) \\
            $2 \leq p \leq k-2$
        \end{tabular}  & 
\begin{tikzpicture}
\node[circle,fill=black,inner sep=0pt,minimum size=10pt,label=below:{$0$}] (0) at (-.707*\x,.707*\x) {};
\node[circle,draw=black,inner sep=0pt,minimum size=10pt,label=below:{$1$}] (1) at (-.707*\x,-.707*\x) {};
\node[circle,fill=black,inner sep=0pt,minimum size=10pt,label=below:{$2$}] (2) at (0*\x,0*\x) {};
\node[circle,fill=black,inner sep=0pt,minimum size=10pt,label=below:{$3$}] (3) at (1*\x,0*\x) {};
\node (4) at (2*\x,0*\x) {...};
\node[circle,fill=black,inner sep=0pt,minimum size=10pt,label=below:{$p$}] (5) at (3*\x,0*\x) {};
\node[circle,draw=black,inner sep=0pt,minimum size=10pt,label=below:{$p+1$}] (6) at (4*\x,0*\x) {};
\node (7) at (5*\x,0*\x) {...};
\node[circle,draw=black,inner sep=0pt,minimum size=10pt,label=below:{$k-1$}] (8) at (6*\x,0) {};
\node[circle,draw=black,inner sep=0pt,minimum size=10pt,label=below:{$k$}] (9) at (7*\x,0) {};
\draw[-] (0) -- (2);
\draw[-] (1) -- (2);
\draw[-] (2) -- (3);
\draw[-] (3) -- (4);
\draw[-] (5) -- (4);
\draw[-] (5) -- (6);
\draw[-] (6) -- (7);
\draw[-] (8) -- (7);
\draw[transform canvas={yshift=2*\x}] (9) -- (8);
\draw[transform canvas={yshift=-2*\x}] (9) -- (8);
\draw (-.707*\x,.707*\x) circle (.3*\x);
\draw[-] (6.6*\x,0*\x) -- (6.4*\x,.2*\x);
\draw[-] (6.6*\x,0*\x) -- (6.4*\x,-.2*\x);
\end{tikzpicture}\\ \hline
       (I,$k-1$) & 
\begin{tikzpicture}
\node[circle,fill=black,inner sep=0pt,minimum size=10pt,label=below:{$0$}] (0) at (-.707*\x,.707*\x) {};
\node[circle,draw=black,inner sep=0pt,minimum size=10pt,label=below:{$1$}] (1) at (-.707*\x,-.707*\x) {};
\node[circle,fill=black,inner sep=0pt,minimum size=10pt,label=below:{$2$}] (2) at (0*\x,0*\x) {};
\node[circle,fill=black,inner sep=0pt,minimum size=10pt,label=below:{$3$}] (3) at (1*\x,0*\x) {};
\node (4) at (2*\x,0*\x) {...};
\node[circle,fill=black,inner sep=0pt,minimum size=10pt,label=below:{$p$}] (5) at (3*\x,0*\x) {};
\node[circle,fill=black,inner sep=0pt,minimum size=10pt,label=below:{$p+1$}] (6) at (4*\x,0*\x) {};
\node (7) at (5*\x,0*\x) {...};
\node[circle,fill=black,inner sep=0pt,minimum size=10pt,label=below:{$k-1$}] (8) at (6*\x,0) {};
\node[circle,draw=black,inner sep=0pt,minimum size=10pt,label=below:{$k$}] (9) at (7*\x,0) {};
\draw[-] (0) -- (2);
\draw[-] (1) -- (2);
\draw[-] (2) -- (3);
\draw[-] (3) -- (4);
\draw[-] (5) -- (4);
\draw[-] (5) -- (6);
\draw[-] (6) -- (7);
\draw[-] (8) -- (7);
\draw[transform canvas={yshift=2*\x}] (9) -- (8);
\draw[transform canvas={yshift=-2*\x}] (9) -- (8);
\draw (-.707*\x,.707*\x) circle (.3*\x);
\draw[-] (6.6*\x,0*\x) -- (6.4*\x,.2*\x);
\draw[-] (6.6*\x,0*\x) -- (6.4*\x,-.2*\x);
\end{tikzpicture} \\ \hline
      II  & \begin{tikzpicture}
\node[circle,fill=black,inner sep=0pt,minimum size=10pt,label=below:{$0$}] (0) at (-.707*\x,.707*\x) {};
\node[circle,draw=black,inner sep=0pt,minimum size=10pt,label=below:{$1$}] (1) at (-.707*\x,-.707*\x) {};
\node[circle,fill=black,inner sep=0pt,minimum size=10pt,label=below:{$2$}] (2) at (0*\x,0*\x) {};
\node[circle,fill=black,inner sep=0pt,minimum size=10pt,label=below:{$3$}] (3) at (1*\x,0*\x) {};
\node (4) at (2*\x,0*\x) {...};
\node[circle,fill=black,inner sep=0pt,minimum size=10pt,label=below:{$p$}] (5) at (3*\x,0*\x) {};
\node[circle,fill=black,inner sep=0pt,minimum size=10pt,label=below:{$p+1$}] (6) at (4*\x,0*\x) {};
\node (7) at (5*\x,0*\x) {...};
\node[circle,fill=black,inner sep=0pt,minimum size=10pt,label=below:{$k-1$}] (8) at (6*\x,0) {};
\node[circle,fill=black,inner sep=0pt,minimum size=10pt,label=below:{$k$}] (9) at (7*\x,0) {};
\draw[-] (0) -- (2);
\draw[-] (1) -- (2);
\draw[-] (2) -- (3);
\draw[-] (3) -- (4);
\draw[-] (5) -- (4);
\draw[-] (5) -- (6);
\draw[-] (6) -- (7);
\draw[-] (8) -- (7);
\draw[transform canvas={yshift=2*\x}] (9) -- (8);
\draw[transform canvas={yshift=-2*\x}] (9) -- (8);
\draw (-.707*\x,.707*\x) circle (.3*\x);
\draw[-] (6.6*\x,0*\x) -- (6.4*\x,.2*\x);
\draw[-] (6.6*\x,0*\x) -- (6.4*\x,-.2*\x);
\end{tikzpicture}\\ \hline
       III & 
\begin{tikzpicture}
\node[circle,fill=black,inner sep=0pt,minimum size=10pt,label=below:{$0$}] (0) at (-.707*\x,.707*\x) {};
\node[circle,draw=black,inner sep=0pt,minimum size=10pt,label=below:{$1$}] (1) at (-.707*\x,-.707*\x) {};
\node[circle,fill=black,inner sep=0pt,minimum size=10pt,label=below:{$2$}] (2) at (0*\x,0*\x) {};
\node[circle,fill=black,inner sep=0pt,minimum size=10pt,label=below:{$3$}] (3) at (1*\x,0*\x) {};
\node (4) at (2*\x,0*\x) {...};
\node[circle,fill=black,inner sep=0pt,minimum size=10pt,label=below:{$p$}] (5) at (3*\x,0*\x) {};
\node[circle,fill=black,inner sep=0pt,minimum size=10pt,label=below:{$p+1$}] (6) at (4*\x,0*\x) {};
\node (7) at (5*\x,0*\x) {...};
\node[circle,fill=black,inner sep=0pt,minimum size=10pt,label=below:{$k-1$}] (8) at (6*\x,0) {};
\node[circle,draw=black,inner sep=0pt,minimum size=10pt,label=below:{$k$}] (9) at (7*\x,0) {};
\draw[-] (0) -- (2);
\draw[-] (1) -- (2);
\draw[-] (2) -- (3);
\draw[-] (3) -- (4);
\draw[-] (5) -- (4);
\draw[-] (5) -- (6);
\draw[-] (6) -- (7);
\draw[-] (8) -- (7);
\draw[transform canvas={yshift=2*\x}] (9) -- (8);
\draw[transform canvas={yshift=-2*\x}] (9) -- (8);
\draw (-.707*\x,.707*\x) circle (.3*\x);
\filldraw[fill=black, draw=black] (7*\x,0) circle (.1*\x);
\draw[-] (6.6*\x,0*\x) -- (6.4*\x,.2*\x);
\draw[-] (6.6*\x,0*\x) -- (6.4*\x,-.2*\x);
\end{tikzpicture} \\ \hline
    \end{tabular}
    \caption{Root system types for $\mathfrak{so}(2,2k-1)$ (for $k \geq 3$). The circled node 0 is the maximal
    non-compact root $\beta$, equal to the affine root. The non-circled black nodes are the roots that are orthogonal to $\Lambda$. Because of the constraints on $\Lambda$, the root 1 can never be orthogonal to $\Lambda$.  
    The small black dot means that the root satisfies $\langle \Lambda , \alpha \rangle = \frac{1}{2}$. The root system $Q(\Lambda)$ is the one generated by the big black dots, and the root system $R(\Lambda)$ is the one generated by all the black dots.  }
    \label{tabUnitaryBk}
\end{table}

\begin{table}[]
    \centering
    \begin{tabular}{|c|c|c|}
    \hline
        Type & Definition & Unitarity constraint \\ \hline
        (I,$p$) &  
        \begin{tabular}{cc}
            $\lambda_2 = \dots = \lambda_{p+1} > \lambda_{p+2}$ & ($1 \leq p \leq k-2$) \\
            $\lambda_2 = \dots = \lambda_{k} \notin \{0,\frac{1}{2}\}$ & ($p=k-1$)
        \end{tabular}
        & $ E \geq 2k-2+ \lambda_2-p$ \\ \hline 
          II &  
        $\lambda_2 = \dots = \lambda_{k}=0$
        & $ E \geq k-\frac{3}{2}$ or $E=0$\\ \hline   
                 III &  
        $\lambda_2 = \dots = \lambda_{k}=1/2$
        & $ E \geq k-1$ \\ \hline    
    \end{tabular}
    \caption{Unitarity conditions for $\mathfrak{so}(2,2k-1)$. Here the $(\lambda_2 , \dots , \lambda_k)$ are either all integers or all half-integers, and $\lambda_2 \geq \dots \geq \lambda_k \geq 0$. We use interchangeably the notation $E=-\lambda_1$.  }
    \label{TableUnitarityBk}
\end{table}
\clearpage
}

\subsection{\texorpdfstring{The Algebras $D_k=\mathfrak{so}(2,2k-2)$}{}}
\label{unitaryeven}
In this subsection, we list  the unitary highest weight modules of the conformal algebra in
even dimensions. We again present highlights of the classification theorem proven in detail in  \cite{enright1983classification}. The final result in all even dimensions can also be summarized very succinctly.
\afterpage{
\begin{table}[]
\def\x{1}
    \centering
    \begin{tabular}{|c|c|}
    \hline
       (I,1)  & 
\begin{tikzpicture}
\node[circle,fill=black,inner sep=0pt,minimum size=10pt,label=below:{$0$}] (0) at (-.707*\x,.707*\x) {};
\node[circle,draw=black,inner sep=0pt,minimum size=10pt,label=below:{$1$}] (1) at (-.707*\x,-.707*\x) {};
\node[circle,draw=black,inner sep=0pt,minimum size=10pt,label=below:{$2$}] (2) at (0*\x,0*\x) {};
\node (3) at (1*\x,0*\x) {...};
\node[circle,draw=black,inner sep=0pt,minimum size=10pt,label=below:{$p$}] (4) at (2*\x,0*\x) {};
\node[circle,draw=black,inner sep=0pt,minimum size=10pt,label=below:{$p+1$}] (5) at (3*\x,0*\x) {};
\node (6) at (4*\x,0*\x) {...};
\node[circle,draw=black,inner sep=0pt,minimum size=10pt,label=right:{$k-2$}] (7) at (5*\x,0) {};
\node[circle,draw=black,inner sep=0pt,minimum size=10pt,label=right:{$k-1$}] (8) at (5.707*\x,.707*\x) {};
\node[circle,draw=black,inner sep=0pt,minimum size=10pt,label=right:{$k$}] (9) at (5.707*\x,-.707*\x) {};
\draw[-] (0) -- (2);
\draw[-] (1) -- (2);
\draw[-] (2) -- (3);
\draw[-] (3) -- (4);
\draw[-] (4) -- (5);
\draw[-] (5) -- (6);
\draw[-] (6) -- (7);
\draw[-] (7) -- (8);
\draw[-] (7) -- (9);
\draw (-.707*\x,.707*\x) circle (.3*\x);
\end{tikzpicture}\\ \hline
          \begin{tabular}{c}
           (I,$p$) \\
            $2 \leq p \leq k-2$
        \end{tabular}  & \begin{tikzpicture}
\node[circle,fill=black,inner sep=0pt,minimum size=10pt,label=below:{$0$}] (0) at (-.707*\x,.707*\x) {};
\node[circle,draw=black,inner sep=0pt,minimum size=10pt,label=below:{$1$}] (1) at (-.707*\x,-.707*\x) {};
\node[circle,fill=black,inner sep=0pt,minimum size=10pt,label=below:{$2$}] (2) at (0*\x,0*\x) {};
\node (3) at (1*\x,0*\x) {...};
\node[circle,fill=black,inner sep=0pt,minimum size=10pt,label=below:{$p$}] (4) at (2*\x,0*\x) {};
\node[circle,draw=black,inner sep=0pt,minimum size=10pt,label=below:{$p+1$}] (5) at (3*\x,0*\x) {};
\node (6) at (4*\x,0*\x) {...};
\node[circle,draw=black,inner sep=0pt,minimum size=10pt,label=right:{$k-2$}] (7) at (5*\x,0) {};
\node[circle,draw=black,inner sep=0pt,minimum size=10pt,label=right:{$k-1$}] (8) at (5.707*\x,.707*\x) {};
\node[circle,draw=black,inner sep=0pt,minimum size=10pt,label=right:{$k$}] (9) at (5.707*\x,-.707*\x) {};
\draw[-] (0) -- (2);
\draw[-] (1) -- (2);
\draw[-] (2) -- (3);
\draw[-] (3) -- (4);
\draw[-] (4) -- (5);
\draw[-] (5) -- (6);
\draw[-] (6) -- (7);
\draw[-] (7) -- (8);
\draw[-] (7) -- (9);
\draw (-.707*\x,.707*\x) circle (.3*\x);
\end{tikzpicture} \\ \hline
       (I,$k-1$) & \begin{tikzpicture}
\node[circle,fill=black,inner sep=0pt,minimum size=10pt,label=below:{$0$}] (0) at (-.707*\x,.707*\x) {};
\node[circle,draw=black,inner sep=0pt,minimum size=10pt,label=below:{$1$}] (1) at (-.707*\x,-.707*\x) {};
\node[circle,fill=black,inner sep=0pt,minimum size=10pt,label=below:{$2$}] (2) at (0*\x,0*\x) {};
\node (3) at (1*\x,0*\x) {...};
\node[circle,fill=black,inner sep=0pt,minimum size=10pt,label=below:{$p$}] (4) at (2*\x,0*\x) {};
\node[circle,fill=black,inner sep=0pt,minimum size=10pt,label=below:{$p+1$}] (5) at (3*\x,0*\x) {};
\node (6) at (4*\x,0*\x) {...};
\node[circle,fill=black,inner sep=0pt,minimum size=10pt,label=right:{$k-2$}] (7) at (5*\x,0) {};
\node[circle,fill=black,inner sep=0pt,minimum size=10pt,label=right:{$k-1$}] (8) at (5.707*\x,.707*\x) {};
\node[circle,draw=black,inner sep=0pt,minimum size=10pt,label=right:{$k$}] (9) at (5.707*\x,-.707*\x) {};
\draw[-] (0) -- (2);
\draw[-] (1) -- (2);
\draw[-] (2) -- (3);
\draw[-] (3) -- (4);
\draw[-] (4) -- (5);
\draw[-] (5) -- (6);
\draw[-] (6) -- (7);
\draw[-] (7) -- (8);
\draw[-] (7) -- (9);
\draw (-.707*\x,.707*\x) circle (.3*\x);
\end{tikzpicture}\\ \hline
       (I,$k-1$)' & \begin{tikzpicture}
\node[circle,fill=black,inner sep=0pt,minimum size=10pt,label=below:{$0$}] (0) at (-.707*\x,.707*\x) {};
\node[circle,draw=black,inner sep=0pt,minimum size=10pt,label=below:{$1$}] (1) at (-.707*\x,-.707*\x) {};
\node[circle,fill=black,inner sep=0pt,minimum size=10pt,label=below:{$2$}] (2) at (0*\x,0*\x) {};
\node (3) at (1*\x,0*\x) {...};
\node[circle,fill=black,inner sep=0pt,minimum size=10pt,label=below:{$p$}] (4) at (2*\x,0*\x) {};
\node[circle,fill=black,inner sep=0pt,minimum size=10pt,label=below:{$p+1$}] (5) at (3*\x,0*\x) {};
\node (6) at (4*\x,0*\x) {...};
\node[circle,fill=black,inner sep=0pt,minimum size=10pt,label=right:{$k-2$}] (7) at (5*\x,0) {};
\node[circle,draw=black,inner sep=0pt,minimum size=10pt,label=right:{$k-1$}] (8) at (5.707*\x,.707*\x) {};
\node[circle,fill=black,inner sep=0pt,minimum size=10pt,label=right:{$k$}] (9) at (5.707*\x,-.707*\x) {};
\draw[-] (0) -- (2);
\draw[-] (1) -- (2);
\draw[-] (2) -- (3);
\draw[-] (3) -- (4);
\draw[-] (4) -- (5);
\draw[-] (5) -- (6);
\draw[-] (6) -- (7);
\draw[-] (7) -- (8);
\draw[-] (7) -- (9);
\draw (-.707*\x,.707*\x) circle (.3*\x);
\end{tikzpicture}\\ \hline
      II  & \begin{tikzpicture}
\node[circle,fill=black,inner sep=0pt,minimum size=10pt,label=below:{$0$}] (0) at (-.707*\x,.707*\x) {};
\node[circle,draw=black,inner sep=0pt,minimum size=10pt,label=below:{$1$}] (1) at (-.707*\x,-.707*\x) {};
\node[circle,fill=black,inner sep=0pt,minimum size=10pt,label=below:{$2$}] (2) at (0*\x,0*\x) {};
\node (3) at (1*\x,0*\x) {...};
\node[circle,fill=black,inner sep=0pt,minimum size=10pt,label=below:{$p$}] (4) at (2*\x,0*\x) {};
\node[circle,fill=black,inner sep=0pt,minimum size=10pt,label=below:{$p+1$}] (5) at (3*\x,0*\x) {};
\node (6) at (4*\x,0*\x) {...};
\node[circle,fill=black,inner sep=0pt,minimum size=10pt,label=right:{$k-2$}] (7) at (5*\x,0) {};
\node[circle,fill=black,inner sep=0pt,minimum size=10pt,label=right:{$k-1$}] (8) at (5.707*\x,.707*\x) {};
\node[circle,fill=black,inner sep=0pt,minimum size=10pt,label=right:{$k$}] (9) at (5.707*\x,-.707*\x) {};
\draw[-] (0) -- (2);
\draw[-] (1) -- (2);
\draw[-] (2) -- (3);
\draw[-] (3) -- (4);
\draw[-] (4) -- (5);
\draw[-] (5) -- (6);
\draw[-] (6) -- (7);
\draw[-] (7) -- (8);
\draw[-] (7) -- (9);
\draw (-.707*\x,.707*\x) circle (.3*\x);
\end{tikzpicture} \\ \hline
    \end{tabular}
    \caption{Root system types for $\mathfrak{so}(2,2k-2)$ (with $k \geq 2$). The circled node 0 is the affine root $\beta$. The non-circled black nodes are the roots that are orthogonal to $\Lambda$. Because of the conventions for $\Lambda$, the root 1 can never be orthogonal to $\Lambda$.  }
       \label{tabUnitaryDk}
\end{table}
\begin{table}[]
    \centering
        \begin{tabular}{|c|c|c|}
    \hline
        Type & Definition & Unitarity constraint \\ \hline
        (I,$p$) &  
        \begin{tabular}{cc}
            $\lambda_2 = \dots = \lambda_{p+1} > |\lambda_{p+2}|$ & ($1 \leq p \leq k-2$) \\
            $\lambda_2 = \dots = \lambda_{k} \neq 0$ & for (I,$k-1$) \\
            $\lambda_2 = \dots = -\lambda_{k} \neq 0$ & for (I,$k-1$)' 
        \end{tabular}
        & $ E \geq 2k-3+ \lambda_2-p$ \\ \hline 
          II &  
        $\lambda_2 = \dots = \lambda_{k}=0$
        & $ E \geq k-2$ or $E=0$\\ \hline   
    \end{tabular}
    \caption{Unitarity conditions for $\mathfrak{so}(2,2k-2)$. Here $\lambda_2 \geq \dots \geq \lambda_{k-1} \geq |\lambda_k|$, all the differences $\lambda_i - \lambda_{i+1} \in \mathbb{Z}$ for $i=2, \dots , k-1$ and $2 \lambda_k \in \mathbb{Z}$. We use interchangeably the notation $E=-\lambda_1$.  }
     \label{TableUnitarityDk}
\end{table}
\clearpage
}
The ground work is layed by noting that the set of compact positive roots $\Phi_c^+$ and
the set of non-compact positive roots $\Phi_n^+$ is given for the $D_k$ algebra by
\begin{eqnarray}
\Phi_c^+ &=& \{ e_i \pm e_j | 2 \le i < j \le k \}
\nonumber \\
\Phi_n^+ &=& \{ e_1 \pm e_j | 2 \le j \le k \} \, .
\end{eqnarray}
The highest non-compact root $\beta$ again coincides with the highest root for this algebra 
\begin{equation}
\beta = e_1+e_2 \, ,
\end{equation}
and the Weyl vector is 
\begin{equation}
\rho =(k-1,k-2,\dots,0) \, .
\end{equation}
We use the following parameterization of the anchor weight $\Lambda$ and the orthogonal pointer weight 
$\zeta$
\begin{eqnarray}
\Lambda &=& (\lambda_1,\dots, \lambda_n)
\nonumber \\
\zeta &=& (1,0,\dots,0) \, .
\end{eqnarray}
The anchor weight $\Lambda$ is dominant in its compact components:
\begin{eqnarray}
 \lambda_2 \ge \lambda_3 \ge \dots \ge \lambda_{k-1} \ge |\lambda_k| & &
\nonumber \\
 \lambda_i-\lambda_j \in \mathbb{N} && (2 \le i < j \le k)
\nonumber \\
2 \lambda_k \in \mathbb{Z}& &
\nonumber \\
\lambda_1 + \lambda_2 = -2k+3 \, . &&
\end{eqnarray}
The root systems $Q(\Lambda)$ and $R(\Lambda)$ are always equal since all roots have equal length.
There are again three cases to distinguish, but two of them are related by the outer
automorphism of $\mathfrak{so}(2,2k-2)$. The latter acts on the weight components by flipping the sign
of the final component $\lambda_k$. This symmetry of our classification problem reduces the number of cases
to two, namely the root systems 
$\mathfrak{su}(1,p)$ with $p \le k-1$, and the root system $\mathfrak{so}(2,2k-2)$.
The final statement is that
the representation is unitarizable if and only if 
\begin{eqnarray}
z \le p  & &  \mbox{ for case (I,p)}
\nonumber \\
z \le k-1  &&  \mbox{ for case II} \, ,
\end{eqnarray}
with the exception of $z=2k-3$ in case II, which corresponds to the trivial
representation of $\mathfrak{so}(2k-2)$.  We now determine in which case we are, 
depending on the weight $\Lambda$.

 The root system $Q(\Lambda)$ is simple, and
contains at least the maximal non-compact root $\beta=e_1+e_2$. Thus, we consider first whether
the compact roots containing an $e_2$ term belong to the root system $Q(\Lambda)$. Given the constraints on the weights $\lambda_i$ of the finite dimensional representation of $\mathfrak{so}(2k-2)$, this is the case if and only if
$\lambda_2=\lambda_3$. If these entries are not equal, then the root system $Q(\Lambda)$ corresponds to the rank one non-compact algebra $\mathfrak{su}(1,1)$. When $\lambda_2=\lambda_3$, we attach one further node. We continue
in this manner, and find that when we have consecutive components $\lambda_2=\lambda_3=\dots=\lambda_p=\lambda_{p+1}$
equal, then the non-compact algebra is $\mathfrak{su}(1,p)$. When we reach the end of the chain, we have the case
$\lambda_2=|\lambda_k|\neq 0$ with algebra $\mathfrak{su}(1,k-1)$. Finally, we have the exceptional 
case 
$\lambda_2=0$ for which the root system $Q(\Lambda)$ corresponds to the full algebra $\mathfrak{so}(2,2k-2)$. Thus, for each weight $\Lambda$, we have found the root system $Q(\Lambda)$. We can then summarize all unitary highest weight representations. 
We again declare $\lambda_1 = -E$, and we run through all possible cases. We list the results for $\mathfrak{so}(2,2k-2)$ in Tables \ref{tabUnitaryDk} and \ref{TableUnitarityDk}.

\section{The Weyl Group Cosets}
\label{factorization}

In this section, we combine the results of sections \ref{characters} and \ref{unitary} to compute the characters of the unitary irreducible representations of the conformal algebras in various dimensions.
We already saw how the generic representation theory boils down to Weyl group theory and the calculation
of Kazhdan-Lusztig polynomials (evaluated at one). We will further show that in the case of 
unitary representations, the Weyl group combinatorics can be simplified by performing an efficient (parabolic) decomposition. The paper \cite{Penedones:2015aga} summarizes some of the features
that we exhibit in detail. See also \cite{Beccaria:2014jxa,Basile:2016aen} for a related approach, based on the exact sequences of \cite{Shaynkman:2004vu}.

Consider the conformal algebra  $\mathfrak{so}(2,n)$ in $n$ dimensions and its Weyl group $W$. The Weyl group is generated by the simple reflections $S = \{s_1 , \dots , s_{n+1}\}$. The Weyl group of the compact subalgebra $\mathfrak{so}(n)$ is generated by the reflections $J = \{s_2 , \dots , s_{n+1}\}$. We call this Weyl group $W_J$. Finally, we construct the set $W^J \subset W$ by taking, in each equivalence class of $W_J \backslash W$, the element of minimal length. Then, each $w \in W$ can be written in a unique way as
\begin{equation}
    w = w_J w^J \, , \qquad w_J \in W_J \, , w^J \in W^J \, . 
\end{equation}
In particular, the longest element $w_{\circ}$ of $W$ decomposes as $w_{\circ,J} w^J_{\circ}$, where $w_{\circ,J}$ is the longest element of $W_J$.

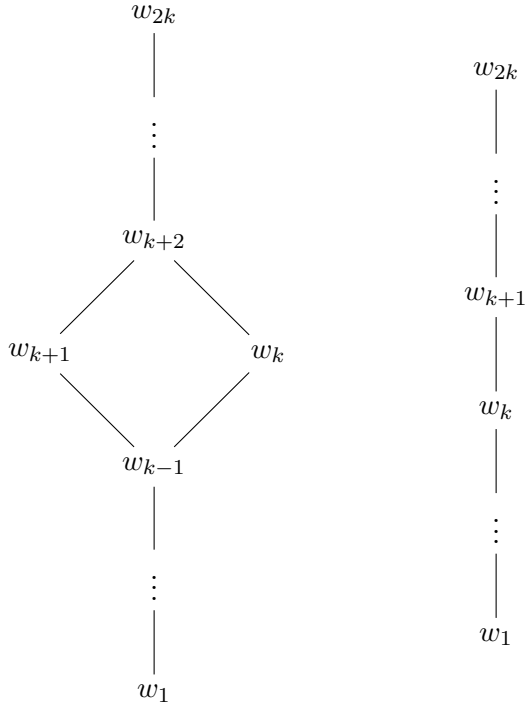
\begin{figure}[t]
    \centering
    \def\x{1.5}
    \def\y{1.5}
 \begin{tikzpicture}
\node[] (1) at (0.*\x,0*\y){$w_1$};
\node (2) at (0.*\x,1*\y){$\vdots$};
\node (3) at (0.*\x,2*\y){$w_{k-1}$};
\node (4) at (1.*\x,3*\y){$w_{k}$};
\node (5) at (-1.*\x,3*\y){$w_{k+1}$};
\node (6) at (0.*\x,4*\y){$w_{k+2}$};
\node (7) at (0.*\x,5*\y){$\vdots$};
\node (8) at (0.*\x,6*\y){$w_{2k}$};
\draw[-] (2)-- (1);
\draw[-] (2)-- (3);
\draw[-] (3)-- (4);
\draw[-] (3)-- (5);
\draw[-] (4)-- (6);
\draw[-] (5)-- (6);
\draw[-] (6)-- (7);
\draw[-] (7)-- (8);
\node (11) at (3.*\x,.5*\y){$w_1$};
\node (12) at (3.*\x,1.5*\y){$\vdots$};
\node (13) at (3.*\x,2.5*\y){$w_{k}$};
\node (14) at (3.*\x,3.5*\y){$w_{k+1}$};
\node (15) at (3.*\x,4.5*\y){$\vdots$};
\node (16) at (3.*\x,5.5*\y){$w_{2k}$};
\draw[-] (11)-- (12);
\draw[-] (12)-- (13);
\draw[-] (13)-- (14);
\draw[-] (14)-- (15);
\draw[-] (15)-- (16);
\end{tikzpicture}   
    \caption{Bruhat order for $W^J$ for $D_k$ (left) and $B_k$ (right).  }
    \label{figBruhatWJBD}
\end{figure}

The highest weights of unitary irreducible representations of the conformal algebras are dominant in the compact direction. Let $\lambda = w \cdot \bl$ be such a weight, with $w \in W$ and $\bl$ antidominant. Because $\lambda$ is dominant in the compact direction, the parabolic decomposition of $w$ reads $w = w_{\circ,J} w^J$. 
Let $v=v_J v^J \in W$. We want to evaluate $P_{v,w}$. We have that \cite{Deodhar}\footnote{In the last step, following Proposition 3.4 in \cite{Deodhar}, we have introduced the standard notation for
parabolic KL polynomials. In the context of our paper, 
we can take the last equation as their definition. Since the notation 
matches the mathematics literature, comparison is easier.
} 
\begin{equation}
\label{KLparab}
    P_{v,w} = P_{v_J v^J , w_{\circ,J} w^J} = P_{w_{\circ,J} v^J , w_{\circ,J} w^J} = P^{J,-1}_{v^J,w^J} \, . 
\end{equation}
We see that for an element $w$ of the form $w = w_{\circ,J} w^J$, the polynomial $P_{v,w}$ depends only on the representatives of $v$ and $w$ in $W^J$. 
For that reason, it is necessary to study the structure of $W^J$. 
The Bruhat order in $W^J$ is given in Figure \ref{figBruhatWJBD} for the conformal
algebras. Using the notations of the Figure, we have that \cite{boe1988kazhdan,brenti2009parabolic}
\begin{itemize}
    \item For $B_n$, 
    \begin{equation}
    \label{parabKLB}
        P^{J,-1}_{w_i,w_j} = \begin{cases}
        1 & i \leq j \\
        0 & \textrm{otherwise}. 
        \end{cases}
    \end{equation}
    \item For $D_n$,    
    \begin{equation}
    \label{parabKLD}
        P^{J,-1}_{w_i,w_j} = \begin{cases}
        0 & w_i \nleq w_j \\
        1+q^{j-n-1} & n+2 \leq j \leq 2n-1 \, , 1 \leq i \leq 2n-j \\
        1 & \textrm{otherwise} \, . 
        \end{cases}
    \end{equation}
\end{itemize}
We note the drastic simplification in the complexity of the Kazhdan-Lusztig polynomials.
They can be explicitly computed, and when evaluated at one, they equal no more than two.

These preliminaries allow us to simplify the character formula.
Let $\lambda$ be a unitary weight. It can be written 
\begin{equation}
\label{notationLambdac}
    \lambda = (\lambda_1  , \underbrace{\lambda_2 , \dots , \lambda_k}_{\lambda^c})
\end{equation}
where $\lambda^c$ is a dominant integral weight of the compact subalgebra. As such, it is the highest weight of a finite dimensional representation of the compact algebra $\mathfrak{k}$, whose character is denoted by $[L^{\mathfrak{k}}_{\lambda}]$. This generalizes equation (\ref{su2finite}). Then we can introduce the generalized Verma modules $M^c_{\lambda}$, defined from those finite-dimensional representations of the compact subalgebra by induction to the full algebra. The characters are related by 
\begin{equation}
    [M^c_{\lambda}] = \frac{e^{\lambda_1}}{\prod\limits_{\alpha \in \Phi_n^+} (1-z^{- \alpha})} [L^{\mathfrak{k}}_{\lambda}] \label{inducedVerma}
\end{equation}
and the Weyl character formula gives 
\begin{equation}
    [M^c_{\lambda}] = \sum\limits_{w \in W_J} (-1)^{\ell (w)} [M_{w \cdot \lambda}] \, . 
\end{equation}
Our goal is now to express the irreducible characters $[L_{\lambda}]$ in terms of the 
induced characters $ [M^c_{\mu}] $. We have, for $\lambda$ integral and unitary, 
\begin{eqnarray}
    \left[L_{w \cdot \bl}\right] &=& \sum\limits_{w' \in W} (-1)^{\ell(w,w')} P_{w',w}(1) \left[M_{w' \cdot \bl}\right] \\ \nonumber
    &=& \sum\limits_{w'_J \in W_J}  \sum\limits_{w'^J \in W^J} (-1)^{\ell(w_{\circ,J} w^J,w'_Jw'^J)} P_{w'_Jw'^J,w_{\circ,J} w^J}(1) \left[M_{w'_J w'^J \cdot \bl}\right] \\ \nonumber
    &=& \sum\limits_{w'_J \in W_J}  \sum\limits_{w'^J \in W^J} (-1)^{\ell(w_{\circ,J})+\ell( w^J)-\ell(w'_J)-\ell(w'^J)} P_{w_{\circ,J} w'^J,w_{\circ,J} w^J}(1) \left[M_{w'_J w'^J \cdot \bl}\right] \\ \nonumber 
        &=&   \sum\limits_{w'^J \in W^J} (-1)^{\ell(w_{\circ,J})+\ell( w^J)-\ell(w'^J)} P_{w_{\circ,J} w'^J,w_{\circ,J} w^J}(1) \left[M^c_{w_{\circ,J} w'^J \cdot \bl}\right] \, . 
        \label{KLsumWexpJ}
\end{eqnarray}
We have reduced the sum over the Weyl group, which contains respectively $2^{k-1} \, k!$ and $2^k \, k!$ elements for $D_k$ and $B_k$, to a sum over the $2k$ elements of $W^J$.  Thus, unitary highest weight conformal representation theory has been reduced to the analysis of Weyl group parabolic cosets, and their associated Kazhdan-Lusztig polynomials. All the ingredients in the character formula can be explicitly computed.

\section{The  Unitary Conformal Characters}
\label{allunitaries}

In this section, we apply the schemes of sections \ref{characters} and \ref{unitary} to systematically calculate all characters of unitary highest weight representations of the conformal algebras $\mathfrak{so}(2,n)$.

To facilitate the calculations, we give in Tables \ref{tabInequalitiesD} and \ref{tabInequalitiesB} the values of the parabolic coset representatives $w_i$ in terms of simple reflections $s_i$. We number the  simple roots as in the Dynkin diagrams of Tables \ref{tabUnitaryBk} and \ref{tabUnitaryDk}, and denote by $s_i$ the reflection through the simple root $\alpha_i$. Finally, we turn to the longest element\footnote{The longest element of the Weyl group of a simple Lie algebra can be obtained in terms of simple reflections as follows \cite{bourbaki1972groupes,HumphreysCoxeter,59789}. Color the nodes of the Dynkin diagrams in white and black in such a way that no two dots of the same color are connected. Let $w_{\mathrm{black}}$ (respectively $w_{\mathrm{white}}$) be the product of the simple reflections associated to the simple roots painted in black (respectively white). The longest element of the Weyl group is $w_{\circ} = (w_{\mathrm{black}}w_{\mathrm{white}})^{h/2}$ where $h$ is the Coxeter number. Since $-w_{\circ}$ 
induces an automorphism of the Dynkin diagram, we have $w_{\circ} = -1$ for algebras other than $A_k$, $D_k$ and $E_6$. In the case of $D_k$, one finds that $w_{\circ} = -1$ if $k$ is even, and for $k$ odd $-w_{\circ}$ exchanges the last two simple roots. } of $W_J$, which is constructed from the longest element of the subgroups $B_{k-1}$ and $D_{k-1}$. We give them as $k \times k$ matrices
acting on the orthonormal basis $e_i$ of the dual $\mathfrak{h}^{\ast}$ of the Cartan subalgebra, introduced in appendix \ref{conventions}:\footnote{The dots represent minus ones. } 
\begin{equation}
 \label{longestElement}
    w_{\circ,J} = \begin{cases}
    \mathrm{Diag} (+1 , -1 , \dots , -1,-1) & \textrm{ for } B_k \textrm{ and } D_{k \textrm{ odd}} \\
     \mathrm{Diag} (+1 , -1 , \dots , -1 , +1) & \textrm{ for } D_{k \textrm{ even}} \, .  
    \end{cases}
\end{equation}
We first treat odd space-time dimensions, and then even space-time dimensions.

\subsection{In Odd Space-time Dimension}

\begin{table}[]
    \centering
    \begin{tabular}{|c|c|}
      \hline 
                 Element of $W^J$ & 
        \begin{tabular}{c}
            Expression in terms of simple reflections   \\ 
             Inequality
        \end{tabular}
         \\  \hline
        $w_i$ ($1 \leq i \leq k$) & 
        \begin{tabular}{c}
            $w_i = s_1 \dots s_{i-1}$  \\
          $- \lr_2 \leq - \lr_3 \leq \cdots \leq - \lr_i \leq \lr_1 \leq - \lr_{i+1} \leq \cdots \leq - \lr_k \leq 0$
        \end{tabular}
         \\  \hline
        $w_{2k+1-i}$ ($1 \leq i \leq k$) & 
                \begin{tabular}{c}
            $w_{2k+1-i} = s_1 \dots s_{k-1}s_k s_{k-1} \dots s_{i}$  \\
          $- \lr_2 \leq - \lr_3 \leq \cdots \leq - \lr_i \leq - \lr_1 \leq - \lr_{i+1} \leq \cdots \leq - \lr_k \leq 0$
        \end{tabular}
        \\  \hline
    \end{tabular}
    \caption{We write the inequality satisfied by a weight of the form $\lambda = w_{\circ,J} w_i \cdot \bl$ with $\bl$ antidominant.  We recall that $\lambda$ in $B_k$ is antidominant if and only if $\lr_1 \leq \cdots \leq \lr_k \leq 0$.}
    \label{tabInequalitiesB}
\end{table}

Consider the $\mathfrak{so}(2,2k-1)$ conformal algebra. We know that the unitary weights fall into $k+1$ categories (see Table \ref{tabUnitaryBk}). For these unitary weights, we compute the possible Weyl groups $W_{[\lambda]}$. 

There are $k^2$ positive roots: $k$ short roots of the form $\alpha = e_i$, for which $\langle \lambda , \alpha^{\vee} \rangle = 2 \lambda_i$, and $k^2-k$ long roots $e_i \pm e_j$ for $1 \leq i<j \leq k$ for which $\langle \lambda , \alpha^{\vee} \rangle =  \lambda_i \pm \lambda_j$. For a unitary weight, we saw that $(\lambda_2 , \dots , \lambda_k) \in \mathbb{Z}^{k-1} \cup (\mathbb{Z}+ \frac{1}{2})^{k-1}$. Therefore the $(k-1)^2$ roots $e_i \pm e_j$ for $2 \leq i<j \leq k$ and $e_i$ for $2 \leq i \leq k$ satisfy $\langle \lambda , \alpha^{\vee} \rangle \in \mathbb{Z}$. We have to examine the remaining roots $e_1$ and $e_1 \pm e_i$. This leads to the three following possibilities, which define what we call the \emph{integrality class} of the weight $\lambda$: 
\begin{equation}
\label{integralityClassB}
    \begin{array}{|c|c|} 
    \hline
       \textrm{Condition} & W_{[\lambda]}  \textrm{  (Integrality class)} \\ \hline     
        \lambda_1 - \lambda_2 \notin \frac{1}{2} \mathbb{Z} & B_{k-1} \textrm{  (non-integral)}\\ 
       \lambda_1 - \lambda_2 \in \frac{1}{2} + \mathbb{Z} & B_{k-1} \oplus A_1 \textrm{  (half-integral)}\\ 
       \lambda_1 - \lambda_2 \in  \mathbb{Z} &B_{k} \textrm{  (integral)}\\ \hline 
       \end{array}
\end{equation}
We see that in addition to the integral case (where $W_{[\lambda]}=B_k$), there are two other integrality classes to consider, which have  $W_{[\lambda]}=B_{k-1} \oplus A_1$ and $W_{[\lambda]}=B_{k-1}$. We examine them in turn in the following paragraphs. 

In the non-integral case, $W_{[\lambda]}$ reduces to the parabolic Weyl group $W_J$, so we just have $[L_{\lambda}]=[M^c_{\lambda}]$. 

In the half-integral case, we have to take into account a possible reflection with respect to the $A_1$ root, which in our notation is $e_1$. The two $A_1$ Weyl chambers are delimited by the wall $\lambda_1 = \frac{1}{2}-k$. From this, we deduce that 
\begin{itemize}
    \item If $\lambda_1 \leq \frac{1}{2}-k $, then $[L_{\lambda}]=[M^c_{\lambda}]$. 
    \item If $\lambda_1 > \frac{1}{2}-k $, we have to remove a correction, corresponding to the dot image of $\lambda$ under the $e_1$ reflection. Using the notation (\ref{notationLambdac}), this gives
    the character formula
    \begin{equation}
        [L_{\lambda}]=[M^c_{(\lambda_1 , \lambda^c)}] - [M^c_{(1-2k-\lambda_1 , \lambda^c)}] \, . 
    \end{equation}
\end{itemize}

Finally, the integral case is the most complicated one. We apply the Kazhdan-Lusztig formula (\ref{KLgeneric}), where the antidominant weight $\bl$ is needed. If the weight $\lambda$ is singular, several different pairs $(w,\bl)$ can a priori be used in the parameterization (\ref{definitionbl}), but for equation (\ref{KLgeneric}) to be valid, we need to choose the Weyl group element $w$ of minimal length. Moreover we know that $w \in w_{\circ,J} W^J$. In Table \ref{tabInequalitiesB}, we have gathered for each such element $w$ the inequalities that $w \cdot \bl$ satisfies, with $\bl$ antidominant. 
This means that we  select the lowest value of $i$ such that $\lambda$ satisfies the inequality associated to 
the coset representative $w_i$ in Table \ref{tabInequalitiesB}. 
In the table, in order to write the inequalities in a more compact way, we use the shifted notation 
\begin{equation}
    \lambda^{\rho} := \lambda + \rho \, . 
\end{equation}
Then, combining equations (\ref{KLsumWexpJ}) and (\ref{parabKLB}) gives the result. 

\subsection{In Even Space-time Dimension}

\begin{table}[]
    \centering
    \begin{tabular}{|c|c|}
    \hline 
           Element of $W^J$ & 
        \begin{tabular}{c}
            Expression in terms of simple reflections   \\ 
             Inequality
        \end{tabular}
         \\  \hline
        $w_i$ ($1 \leq i \leq k-1$) & 
        \begin{tabular}{c}
            $w_i = s_1 \dots s_{i-1}$   \\ 
              $- \lr_2 \leq \cdots \leq - \lr_i \leq + \lr_1 \leq -\lr_{i+1} \leq \cdots \leq - \lr_{k-1} \leq - |\lr_k|$
        \end{tabular}
         \\  \hline
        $w_k$ & 
         \begin{tabular}{c}
            $w_k = s_1 \dots s_{k-2} s_k $   \\
              $- \lr_2 \leq \cdots \leq - \lr_{k-1} \leq - \lr_{k} \leq - |\lr_1|$
        \end{tabular}
       \\  \hline
        $w_{k+1}$ & 
         \begin{tabular}{c}
          $w_{k+1} = s_1 \dots s_{k-2} s_{k-1} $  \\
             $- \lr_2 \leq \cdots \leq - \lr_{k-1} \leq + \lr_{k} \leq - |\lr_1|$ 
        \end{tabular}
          \\  \hline
        $w_{2k+1-i}$ ($1 \leq i \leq k-1$) & 
         \begin{tabular}{c}
          $w_{2k+1-i} = s_1 \dots  s_{k-1} s_{k} s_{k-2} s_{k-3} \dots s_{i} $   \\
             $- \lr_2 \leq \cdots \leq - \lr_i \leq - \lr_1 \leq -\lr_{i+1} \leq \cdots \leq - \lr_{k-1} \leq - |\lr_k|$
        \end{tabular}
      \\        \hline
    \end{tabular}
    \caption{We write the inequality satisfied by a weight of the form $\lambda = w_{\circ,J} w_j \cdot \bl$ with $\bl$ antidominant. We recall that a weight $\lambda$ in $D_k$ is antidominant if and only if $\lr_1 \leq \cdots \leq \lr_{k-1} \leq - |\lr_k|$.  }
    \label{tabInequalitiesD}
\end{table}

Secondly,  we perform the same analysis for even space-time dimension. The algebra $\mathfrak{so}(2,2k-2)$ possesses $k(k-1)$ positive roots $e_i \pm e_j$ for $1 \leq i < j \leq k$. For a root $\alpha = e_i \pm e_j$, we have $\langle \lambda , \alpha^{\vee} \rangle =  \lambda_i \pm \lambda_j$. A unitary weight has $\lambda_i \pm \lambda_j \in \mathbb{Z}$ for $2 \leq i < j \leq k$, so only two configurations are possible: 
\begin{equation}
\label{integralityClassD}
    \begin{array}{|c|c|} 
    \hline
       \textrm{Condition} & W_{[\lambda]} \textrm{  (Integrality class)} \\ \hline     
        \lambda_1 - \lambda_2 \notin \mathbb{Z} &D_{k-1}\textrm{  (non integral)} \\ 
       \lambda_1 - \lambda_2 \in  \mathbb{Z} &D_{k} \textrm{  (integral)}\\ \hline 
       \end{array}
\end{equation}
In the non-integral case, we  have $[L_{\lambda}]=[M^c_{\lambda}]$. In the integral case, one has to examine the inequalities satisfied by $\lr$. Again, we pick the smallest $w_i$ such that the corresponding inequality in Table \ref{tabInequalitiesD} is satisfied,\footnote{Since the order is only partial, this could be  ill defined if the inequalities of $w_k$ and $w_{k+1}$ could  both be satisfied, and not those for $w_i$, $i \leq k-1$. However, one checks that the inequalities of $w_k$ and $w_{k+1}$ imply $\lr_1=\lr_{k}=0$, and therefore the inequality of $w_{k-1}$ would be satisfied.
Thus, there really is no ambiguity. } and use formula (\ref{KLsumWexpJ}). The compact notation $\lambda^{\rho} : = \lambda + \rho$ is also used. 
Let us study a few explicit examples.

\subsubsection*{Example}
Consider the weight $\lambda=(-1,0,0)$ in $\mathfrak{so}(2,4)$. It is unitary of type II. Moreover, we have $\lr = (1,1,0)$, which satisfies the inequalities for $w_5$ and $w_6$ but not the other $w_i$. So we must write $\lambda = w_{\circ,J} w_5 \cdot \bl$ with $\bl = (-3,-2,0)$. Using the polynomials (\ref{parabKLD}), we obtain
\begin{equation}
    [L_{\lambda}] = [M^c_{w_{\circ,J} w_5 \cdot \bl}] - [M^c_{w_{\circ,J} w_4 \cdot \bl}] - [M^c_{w_{\circ,J} w_3 \cdot \bl}] +   [M^c_{w_{\circ,J} w_2 \cdot \bl}] - 2  [M^c_{w_{\circ,J} w_1 \cdot \bl}] \, . 
\end{equation}
One computes $w_{\circ,J} w_4 \cdot \bl = (-2,0,1)$, $w_{\circ,J} w_3 \cdot \bl = (-2,0,-1)$, and $w_{\circ,J} w_2 \cdot \bl = w_{\circ,J} w_2 \cdot \bl = (-3,0,0)$. The compact part of the weights $(-2,0,\pm 1)$ is singular, so the corresponding module is trivial and disappears in the character formula, as per the remark at the end of section \ref{sectionFiniteDim}. 
We conclude 
\begin{equation}
     [L_{(-1,0,0)}] = [M^c_{(-1,0,0)}] - [M^c_{(-3,0,0)}]  \, . 
\end{equation}
Note that this result arises from cancelling terms that contain multiplicities larger than one.

\subsection*{Example}
Similarly, the weight $\lambda=(-2,0,0,0)$ in $\mathfrak{so}(2,6)$ is associated to the coset
representative $w_6$, and we have 
\begin{equation}
    [L_{\lambda}] = [M^c_{w_{\circ,J} w_6 \cdot \bl}] - [M^c_{w_{\circ,J} w_5 \cdot \bl}] - [M^c_{w_{\circ,J} w_4 \cdot \bl}] +   [M^c_{w_{\circ,J} w_3 \cdot \bl}] - 2  [M^c_{w_{\circ,J} w_2 \cdot \bl}]  + 2  [M^c_{w_{\circ,J} w_1 \cdot \bl}] \, . \nonumber
\end{equation}
This reduces to 
\begin{equation}
     [L_{(-2,0,0,0)}] = [M^c_{(-2,0,0,0)}] - [M^c_{(-4,0,0,0)}]  \, . 
\end{equation}
Again, multiplicities larger than one (and therefore non-trivial Kazhdan-Lusztig polynomials)
play an intermediate role.

\subsubsection*{Conclusion}

We conclude that the calculation of the characters of all highest weight unitary representations of the conformal algebra in any dimension is  straightforward using the mathematical technology. Deciphering suffices.

\section{Summary and Comparison with the Physics Literature}
\label{physics}
A large physics literature exploring the representation theory of the conformal algebras $\mathfrak{so}(2,n)$ is available. The literature concentrates on unitary representations.
These were  classified in three dimensions \cite{Dirac:1963ta,dobrev1991spectrum} and in four dimensions \cite{Mack:1975je}. See also the more general treatment in \cite{Minwalla:1997ka}. The paper
\cite{Ferrara:2000nu} identifies the unitary representations in arbitrary dimensions,
based on the earlier  mathematical treatment in \cite{enright1983classification} which we reviewed in section
\ref{unitary} and which we summarized in  Tables \ref{TableUnitarityBk} and \ref{TableUnitarityDk}. 
Character formulas were computed
in many instances. The most general treatment across dimensions is  \cite{Dolan:2005wy}.\footnote{ However,
the physics literature has not always been entirely accurate, even in the better of resources. See the remark in footnote \ref{VermaMistake}.}

In this section, we translate the uniform mathematical results of section \ref{allunitaries} into a notation  more frequently used by physicists in order to make both the mathematics and the physics literature more  accessible. We again identify the energy $E$, equal to the conformal dimension $\Delta$ of the ground state, with minus the first component of the highest weight, $E=-\lambda_1$. Moreover, the compact subalgebra $\mathfrak{so}(n)$ describes space rotations, and we switch to spin labels $(j_1 , \dots , j_{[n/2]})$
to describe the highest weights of the rotation algebra,\footnote{The spin labels are closer but not yet identical to the most common spin labels in the physics literature. Our convention is uniform across dimensions.}
\begin{equation}
\label{mathPhysTranslation}
    \lambda = \underbrace{(\lambda_1 , \lambda_2 , \dots , \lambda_{1+[n/2]})}_{\textrm{Math}} = \underbrace{(-E , j_1 , \dots , j_{[n/2]})}_{\textrm{Physics}} \, . 
\end{equation} 

\subsection{The Executive Summary}

Firstly though, we summarize the results of sections \ref{unitary}, \ref{factorization} and
\ref{allunitaries} in an effective algorithm that can be used to compute the conformal character -- and indeed the module decomposition -- for any unitary weight. The irreducible conformal character with highest weight $\lambda$ is denoted $[L_{\lambda}]$, and we will obtain an expression in terms of the  Verma modules characters $[M_{\lambda}]$ defined in equation (\ref{definitionM}). 

The procedure runs as follows.
Let $\lambda = (\lambda_1 , \dots , \lambda_k)$ be a weight in $B_k$ or $D_k$.
\begin{enumerate}
    \item Determine whether it is unitary or not using Tables \ref{TableUnitarityBk} and \ref{TableUnitarityDk}. If the weight is not unitary, then the character is given by the general Kazhdan-Lusztig formula (\ref{KLgeneric}). To obtain the character, one needs to compute generic
    Kazhdan-Lusztig polynomials. If the weight is unitary, then a simplification of the
    generic formula occurs, as explained in step two. 
    \item If the weight $\lambda$ is unitary, determine its integrality class 
    using (\ref{integralityClassB}) for a conformal algebra of type $B$, and (\ref{integralityClassD}) for an algebra of type $D$. 
    \begin{itemize}
        \item If the integrality class is $B_{k-1}$ or $D_{k-1}$, then $[L_{\lambda}]=[M^c_{\lambda}]$.
        (See equation (\ref{inducedVerma}) for the character $[M^c_{\lambda}]$ of the Verma module
        induced from an irreducible representation of the compact subalgebra.)
        \item If the integrality class is $B_{k-1} \oplus A_1$, then $[L_{\lambda}]=[M^c_{(\lambda_1,\lambda^c)}] - [M^c_{(1-2k-\lambda_1,\lambda^c)}]$ when $\lambda_1 > \frac{1}{2} -k$, and  $[L_{\lambda}]=[M^c_{\lambda}]$ otherwise. 
        \item If the integrality class is $B_k$ or $D_k$, then look for the \emph{lowest} $w_i$ in Figure \ref{figBruhatWJBD} such that $\lr = \lambda + \rho$ satisfies the
        corresponding inequality in Tables \ref{tabInequalitiesD} and \ref{tabInequalitiesB},\footnote{We recall that $\rho=(k-\frac{1}{2},k-\frac{3}{2},\dots,\frac{1}{2})$ for $B_k$ and $\rho = (k-1,k-2,\dots,0)$ for $D_k$. Moreover, the dot action is defined by $w \cdot \lambda = w(\lambda + \rho) -\rho$. Finally, the longest element of the parabolic Weyl group $w_{\circ,J}$ is given in equation (\ref{longestElement}). } and define $\bl = (w_{\circ,J} w_i)^{-1} \cdot \lambda$. The irreducible character is then given by 
        \begin{equation}
        \label{finalFormlua}
            [L_{\lambda}] = \sum\limits_{j=1}^{2k} (-1)^{\ell(w_i) - \ell(w_j)} b_{ji} [M^c_{w_{\circ,J} w_j \cdot \bl}] \, . 
        \end{equation}
        where the length function $\ell$ is the height in Figure \ref{figBruhatWJBD}, and the multiplicities $b_{ji}$ are obtained by evaluating expressions (\ref{parabKLB}) and (\ref{parabKLD}) at $q=1$, i.e. for $B_k$, 
         \begin{equation}
            \label{parabKLBev}
        b_{ji} = \begin{cases}
        1 & j \leq i \\
        0 & \textrm{otherwise}. 
        \end{cases}
    \end{equation}
    and for $D_k$, 
        \begin{equation}
    \label{parabKLDev}
        b_{ji} = \begin{cases}
        0 & w_j \nleq w_i \\
       2 & k+2 \leq i \leq 2k-1 \, , 1 \leq j \leq 2k-i \\
        1 & \textrm{otherwise} \, . 
        \end{cases}
    \end{equation}
    \end{itemize}
\end{enumerate}

\subsubsection*{Example}
Before we delve into the exhaustive treatment of the low dimensions, we illustrate how the algorithm allows to effectively compute the character of any highest-weight irreducible representation in any dimension. 

Consider the weight $\lambda = (-8,2,2,2,2,1)$ in $\mathfrak{so}(2,10)$. This algebra is of type $D_k$ with $k=6$. First, we check that this weight is unitary. We have $E=- \lambda_1 = 8$, and we observe in Table \ref{TableUnitarityDk} that the unitary constraint is of type (I,4) and reads $E \geq 12 - 3 + 2 - 4 = 7$, which is satisfied. 
To determine the integrality class, we look at Table
(\ref{integralityClassD}), and since $\lambda_1 - \lambda_2 = -10$ is integer, we are in the integral case, called $D_k$.  Hence we are instructed to look in Table \ref{tabInequalitiesD} for the smallest $w_i$, in the order given by Figure \ref{figBruhatWJBD}, such that the corresponding inequality holds. For that, we first compute $\lambda^\rho = (-3,6,5,4,3,1)$. The inequalities for $w_4$ and $w_5$ are both satisfied, but because $w_4$ is smaller than $w_5$, we pick $w_4$.  Now we compute $\bl$. First, note that $w_{\circ,J} = \mathrm{Diag} (1,-1,-1,-1,-1,1)$, and $w_4=s_1 s_2 s_3$ is a cyclic permutation of the four first entries of a weight. So $\bl = (-11,-9,-7,-5,-4,1)$. Finally, the coefficients $b_{ji}$ are non-vanishing only for $j=1,2,3,4$, so we compute the action of $w_{\circ,J} w_j$ for these values of $j$ on $\bl$: $w_{\circ,J} w_1 \cdot \bl = (-11,1,1,1,2,1)$, 
$w_{\circ,J} w_2 \cdot \bl = (-10,2,1,1,2,1)$, $w_{\circ,J} w_1 \cdot \bl = (-9,2,2,1,2,1)$, $w_{\circ,J} w_4 \cdot \bl = (-8,2,2,2,2,1)$. Then, reading the lengths on  Figure \ref{figBruhatWJBD},
we conclude that 
\begin{equation}
    [L_{(-8,2,2,2,2,1)}] = [M^c_{(-8,2,2,2,2,1)}] - [M^c_{(-9,2,2,1,2,1)}] + [M^c_{(-10,2,1,1,2,1)}] - [M^c_{(-11,1,1,1,2,1)}] \, . 
\end{equation} 
In appendix \ref{explicitintegralunitary}, we execute the procedure, and explicitly write down the 
results of formula (\ref{finalFormlua}) for dimensions up to and including seven, for all
integral unitary weights.

\subsection{A Brief Comparison to the Physics Literature}
While the formalism we presented is efficient, it may be beneficial to make an explicit
comparison to results in the literature.
We kick off the brief comparison in three dimensions.

\subsubsection*{\texorpdfstring{Three dimensions: $\mathfrak{so}(2,3)$}{}}
\label{unitaryso23}
We write the highest weights $\lambda = (-E,j)$ in terms of the energy $E$ and spin $j$ of the representation. The unitarity condition of Table \ref{TableUnitarityBk} becomes, with $j \in \frac{1}{2} \mathbb{Z}_{\geq 0}$: 
\begin{itemize}
    \item Type (I,1) is $E \geq j+1$ for $j \neq 0 , \frac{1}{2}$; 
    \item Type II is $E \geq \frac{1}{2}$ or $E=0$ for $j= 0$; 
    \item Type III is $E \geq 1$ for $j=\frac{1}{2}$. 
\end{itemize}
For a unitary weight, we then look at the integrality classes: 
\begin{itemize}
    \item If $2E \notin \mathbb{Z}$, $[L_{\lambda}] = [M^c_{\lambda}]$. 
    \item If $E+j \in \frac{1}{2} + \mathbb{Z}$, then $[L_{(-E,j)}] = [M^c_{(-E,j)}] - [M^c_{(-(3-E),j)}]$ if $E<\frac{3}{2}$, and $[L_{(-E,j)}] = [M^c_{(-E,j)}] $ otherwise. 
    \item Finally, if  $E+j \in \mathbb{Z}$, we look at Table \ref{tabInequalitiesB}, which takes the form 
    \begin{equation}
        \begin{array}{|c|c|}
        \hline 
            w_1 & j \leq E-2   \\ \hline 
            w_2 &  j  \geq E -2 \geq - \frac{1}{2} \\ \hline 
            w_3 & j  \geq -E + 1 \geq -\frac{1}{2} \\ \hline 
            w_4 & j \leq -E+1 \\ \hline 
        \end{array}
    \end{equation}
    We look for the smallest $i$ such that the $w_i$ condition is satisfied by $\lambda$, and read the character in Figure \ref{figBruhatWJBD}. For reference, the results are listed in Table (\ref{caseswiB2}), for each possible value of $i$. In this way we recover the results of section \ref{warmup}. 
    
\end{itemize}

\subsubsection*{\texorpdfstring{Four dimensions: $\mathfrak{so}(2,4)$}{}}
\label{unitaryso24}
In four dimensions, there are three types of unitary weights $\lambda=(-E,j_1,j_2)$.
Table 
\ref{TableUnitarityDk} gives, for $j_1 \geq |j_2|$ and $j_1 - j_2 \in \mathbb{Z}$:\footnote{For comparison with most of the physics literature, one redefines $\tilde{j}_1=(j_1+j_2)/2$
and $\tilde{j}_2=(j_1-j_2)/2$.}
\begin{itemize}
    \item Type (I,1) is $j_1 > |j_2|$ and $E \geq j_1+2$; 
    \item Type (I,2) is $j_1=\pm j_2 \neq 0$ and $E \geq j_1+1$; 
    \item Type II is $j_1=j_2=0$ and $E \geq 1$ or $E=0$. 
\end{itemize}
There are two integrality classes of unitary weights, namely $D_3$ and $D_2$, depending on whether $E-j_1 \in \mathbb{Z}$, or not. In the non integral case, we have $[L_{\lambda}] = [M^c_{\lambda}]$. In the integral case, we look in Table \ref{tabInequalitiesD} for the smallest $i$ such that the appriopriate inequality is satisfied, with $\lambda_1^{\rho} = -E+2$, $\lambda_2^{\rho} = j_1+1$ and $\lambda_2=j_2$. For each value of $i$, the character $[L_{\lambda}]$ can then be read in Table (\ref{caseswiD3}). 
Thus,  we recover the results of \cite{Dolan:2005wy,Barabanschikov:2005ri}.

\subsubsection*{\texorpdfstring{Five dimensions: $\mathfrak{so}(2,5)$}{}}
\label{unitaryso25}

We  distinguish the generic
representations with $E \ge 3 + \lambda_2$ and $\lambda_2>\lambda_3$ (case I,1), the representations
$E \ge 2 + \lambda_2$ where $\lambda_2=\lambda_3$ (case I,2) and the representations with $E \ge 3/2$, which are scalar (case II),
or $E=0$, and $E \ge 2$ for the spinor (case III).
The analysis runs along the lines of the analysis of the conformal algebra $\mathfrak{so}(3,2)$ in three
dimensions. We provide the explicit results for the integral unitary weights in Table (\ref{caseswiB3}). 
When the results can easily be compared, they coincide with \cite{Dolan:2005wy}.
The remark on  singular weights in subsection \ref{sectionFiniteDim} plays a  role
in interpreting the results of \cite{Dolan:2005wy} correctly.

\subsubsection*{\texorpdfstring{Six dimensions: $\mathfrak{so}(2,6)$}{}}
\label{unitaryso26}

The analysis is as for the four dimensional conformal algebra. We provide the explicit results for the integral unitary weights in Table (\ref{caseswiD4}). When the results of \cite{Beccaria:2014qea} can be
unambiguously compared, they agree with ours.

\subsubsection*{Remark on the Generic Case}
Our treatment is generic,
as is \cite{Dolan:2005wy}, but we carefully keep track of possible multiple subtractions of Verma modules.\footnote{Historically, in the mathematical literature, this was not analyzed correctly \cite{humphreys2008representations}. In particular, the otherwise important contribution by Verma \cite{verma1968structure} was mistaken on the possibly larger than one multiplicity of Verma modules to be added or subtracted in the character formula. This has led to wrong claims in the mathematics literature, which unfortunately have propagated to the physics literature (see appendix A of \cite{Dolan:2005wy}). It will be interesting to attempt to prove the character formulas of \cite{Dolan:2005wy}, using the techniques we explained.
\label{VermaMistake}} 
As in \cite{Ferrara:2000nu}, our analysis has the advantage  of being proven necessary and sufficient in arbitrary dimension in regards to unitarity.

\section{Apologia}
Our main aim  was  to provide  physicists with an
overview of the representation theory of conformal multiplets. Highest weight representations make up a large category of representations that is well understood. The minimal data to compute  character formulas for irreducible representations is coded in the Weyl group and the Kazhdan-Lusztig polynomials. 
Mathematicians have also provided a complete analysis of the necessary and sufficient conditions for unitarity, using a more powerful version of the inequalities derived  in the physics literature. Moreover, unitarity restricts the highest weights such that the combinatorial Kazhdan-Lusztig calculations
drastically simplify.

Secondly, by translating mathematics, we have added to the physics literature. We explained how to systematically compute the characters of irreducible highest weight representations even when they are not unitary. We have stressed that the conditions for unitarity are necessary and sufficient, and that they can be formulated at arbitrary rank. In  our analysis, we have dealt systematically with both non-integral as well as singular weights. Moreover, we have provided a clear classification of all cases of unitary characters in terms of coset representatives of a Weyl subgroup of the Weyl group of the conformal algebra. Using our systematic insight, we provided look-up tables for unitary highest weight representation characters for conformal algebras up to and including rank four. They are guaranteed to be complete. Mostly, we hope these tables have gained  in transparency.

Thirdly, these techniques can be refined to apply to superconformal characters. We plan to discuss the necessary extensions elsewhere.

Finally, we wrote this paper because we would have liked to read it.

\section*{Acknowledgements}
We  thank James Humphreys for his writing in general, and his correspondence
in particular. A.B. acknowledges support from the EU CIG grant UE-14-GT5LD2013-618459, the Asturias Government grant FC-15-GRUPIN14-108 and Spanish Government grant MINECO-16-FPA2015-63667-P.
J.T. thanks the High Energy Physics Theory Group of the University of Oviedo for their warm hospitality and acknowledges support from the grant ANR-13-BS05-0001. We thank Xavier Bekaert, Nicolas Boulanger and Hugh Osborn for comments on the first version of this paper.

\appendix

\section{The Lie Algebra Conventions}
\label{conventions}

We use the parameterization of \cite{knapp2013lie} for the roots and weights of the 
$B_k$ and $D_k$ simple Lie algebras. We describe these conventions in detail.
In both cases, the dual of the Cartan subalgebra $\mathfrak{h}^{\ast}$ is spanned by an orthonormal basis $(e_1 , \dots , e_k)$. When we write a weight in components, it is always understood that the coordinates are with respect to this basis. 

\subsection{The Lie Algebra \texorpdfstring{$B_k$}{}}

We have the set of roots
$\Phi=\{ \pm e_i \pm e_j, \pm e_i \}$ and a choice of set of simple roots
$\Phi^s=\{ \alpha_{i<k}=e_i - e_{i+1},\alpha_k=e_k \}$.
The fundamental weights can then be written as
$\varpi_{i<k} =e_1 + \dots + e_i$ and $ \varpi_k=(e_1 + \dots + e_k)/2$. The Weyl vector $\rho$ equals $\rho=((2k-1)e_1+(2k-3)e_2+\dots+ e_k)/2$. The Weyl group is $W_{B_n}=\mathbb{Z}_2^k \rtimes S_k$ and
acts by permutations and sign changes of the orthonormal vectors $e_i$.
The conformal algebra $\mathfrak{so}(2,n)$ with $n$ odd corresponds to a $B_k$ algebra of rank $k=(n+1)/2$.

\subsection{The Lie Algebra \texorpdfstring{$D_k$}{}}
The set of roots is $\Phi=\{ \pm e_i \pm e_j\}$ while simple roots are collected in the set
$\Phi^s=\{ \alpha_{i<k}=e_i - e_{i+1} ,\alpha_k=e_{k-1}+e_k \}$.
The fundamental weights are $\varpi_{i<k-1} =e_1 + \dots + e_i , \varpi_{k-1}=(e_1 + \dots + e_{k-1}-e_k)/2,
\varpi_k=(e_1 + \dots + e_k)/2$. The Weyl vector $\rho$ comes out to be $\rho=(k-1)e_1+(k-2)e_2+\dots+ e_{k-1}$.
The Weyl group is $W_{D_n}=\mathbb{Z}_2^{k-1} \rtimes S_k$ and acts by permuting the vectors $e_i$ 
and
an even number of sign changes.
For the conformal algebra $\mathfrak{so}(2,n)$ with $n$ even, we have a $D_k$ algebra of rank $k=(n+2)/2$.

\section{The Structure of Real Simple Lie Algebras}
\label{realsimple}

We summarize results of the structure theory of semisimple real Lie algebras. We follow
the notation of \cite{knapp2013lie} to which we must refer the reader for a complete exposition.\footnote{Our summary is mainly based on chapters VI on the Structure Theory of 
Semisimple Groups, and chapter VII on the Advanced Structure Theory.}
\subsection{The Structure Theory}
Every complex semisimple algebra $\mathfrak{g}$ has a compact real form. We denote the compact real
forms by $\mathfrak{su}(n),\mathfrak{so}(n),\mathfrak{sp}(n)$ and $\mathfrak{e}_{6,7,8},\mathfrak{f}_4$ and $\mathfrak{g}_2$. The Killing form on a compact semisimple Lie algebra is negative semi-definite
and non-degenerate.

Every real semisimple Lie algebra $\mathfrak{g}_0$ has a Cartan involution $\theta$, unique up to conjugation.
It is such that $B_\theta(X,Y)=-B(X,\theta(Y))$ is positive definite, where $B$ is the Killing form.
This involution $\theta$ gives rise to  an eigenspace decomposition 
\begin{equation}
\mathfrak{g}_0 = \mathfrak{k}_0 \oplus \mathfrak{p}_0
\end{equation}
into eigenspaces of eigenvalues $+1$ and $-1$ respectively. In matrix realizations of Lie algebras,
the Cartan involution $\theta$ can be defined by $\theta (X) = - X^\dagger$, where the dagger stands for the conjugate transpose. The Killing form on $\mathfrak{g}_0$ is negative semi-definite
on $\mathfrak{k}_0$ and positive semi-definite on $\mathfrak{p}_0$.

Every Cartan subalgebra $\mathfrak{h}_0$ of $\mathfrak{g}_0$ is conjugate to a $\theta$-stable subalgebra, and  we will assume that we have picked a Cartan subalgebra
$\mathfrak{h}_0$ that is $\theta$-stable. We can then decompose the Cartan subalgebra into
subalgebras
\begin{equation}
\mathfrak{h}_0 = \mathfrak{t}_0 \oplus \mathfrak{a}_0
\end{equation}
with $\mathfrak{t}_0 \subset \mathfrak{k}_0$ and $\mathfrak{a}_0 \subset \mathfrak{p}_0$. The dimension of $\mathfrak{t}_0$ is called the compact dimension of $\mathfrak{h}_0$, and the dimension of $\mathfrak{a}_0$ is called the non-compact dimension. We say that a Cartan subalgebra is maximally (non-)compact if its (non-)compact dimension is maximal. 

Given a $\theta$-stable Cartan subalgebra $\mathfrak{h}_0 = \mathfrak{t}_0 \oplus \mathfrak{a}_0$, 
the roots of $(\mathfrak{g},\mathfrak{h})$ are imaginary on $\mathfrak{t}_0$ and real on $\mathfrak{a}_0$. As a consequence, we say that a root $\alpha \in \mathfrak{h}_0^\ast$ is {real} if it vanishes on $\mathfrak{t}_0$, and that it is imaginary if it vanishes on $\mathfrak{a}_0$. Otherwise, the root is said to be {complex}. We say that an imaginary root is {compact} if the associated root space is included in $\mathfrak{k}$, and that it is non-compact if it is included in $\mathfrak{p}$. 

To a real semi-simple Lie algebra $\mathfrak{g}_0$, we associate a Vogan diagram which is the 
Dynkin diagram of its complexification $\mathfrak{g}$, adorned with additional data. For a maximally
compact choice of $\mathfrak{h}_0$, there are no real roots. 
Since there are no real roots, we can pick a set of positive roots such that $\theta(\Phi^+)
= \Phi^+$. The Vogan diagram of the triple $(\mathfrak{g}_0,\mathfrak{h}_0,\Phi^+)$ is the Dynkin
diagram of $\Phi^+$ with 2-element orbits of $\theta$ made manifest, and with the 1-element orbits
painted when corresponding to a non-compact simple root, and unpainted when compact \cite{knapp2013lie}.

\subsection{The Classification of Real Simple Lie Algebras}
Firstly, there are the complex simple Lie algebras, considered as an algebra over the real numbers.
Secondly, there are the Lie algebras whose complexification is simple over the complex numbers. These algebras always have a Vogan diagram with at most one simple root painted. Amongst these diagrams,
one can remove further equivalences. The resulting classification of simple real Lie algebras
is summarized e.g. in Theorem 6.105 in \cite{knapp2013lie}. It includes the non-compact forms
$\mathfrak{so}(p,q)$ of the special orthogonal algebras. The Vogan diagram for
$\mathfrak{so}(2,2k-1)$ is 
\begin{equation}
    \def\x{1}
\begin{tikzpicture}
\node[circle,fill=black,inner sep=0pt,minimum size=10pt,label=below:{$1$}] (1) at (0*\x,0*\x) {};
\node[circle,draw=black,inner sep=0pt,minimum size=10pt,label=below:{$2$}] (2) at (1*\x,0*\x) {};
\node (3) at (2*\x,0*\x) {...};
\node[circle,draw=black,inner sep=0pt,minimum size=10pt,label=below:{$k-1$}] (4) at (3*\x,0) {};
\node[circle,draw=black,inner sep=0pt,minimum size=10pt,label=below:{$k$}] (5) at (4*\x,0) {};
\draw[-] (1) -- (2);
\draw[-] (2) -- (3);
\draw[-] (3) -- (4);
\draw[transform canvas={yshift=2*\x}] (4) -- (5);
\draw[transform canvas={yshift=-2*\x}] (4) -- (5);
\draw[-] (3.6*\x,0*\x) -- (3.4*\x,.2*\x);
\draw[-] (3.6*\x,0*\x) -- (3.4*\x,-.2*\x);
\end{tikzpicture}
\end{equation}
and for $\mathfrak{so}(2,2k-2)$, 
\begin{equation}
    \def\x{1}
\begin{tikzpicture}
\node[circle,fill=black,inner sep=0pt,minimum size=10pt,label=below:{$1$}] (1) at (0*\x,0*\x) {};
\node[circle,draw=black,inner sep=0pt,minimum size=10pt,label=below:{$2$}] (2) at (1*\x,0*\x) {};
\node (3) at (2*\x,0*\x) {...};
\node[circle,draw=black,inner sep=0pt,minimum size=10pt,label=right:{$k-2$}] (4) at (3*\x,0) {};
\node[circle,draw=black,inner sep=0pt,minimum size=10pt,label=right:{$k-1$}] (5) at (3.707*\x,.707*\x) {};
\node[circle,draw=black,inner sep=0pt,minimum size=10pt,label=right:{$k$}] (6) at (3.707*\x,-.707*\x) {};
\draw[-] (1) -- (2);
\draw[-] (2) -- (3);
\draw[-] (3) -- (4);
\draw[-] (5) -- (4);
\draw[-] (6) -- (4);
\end{tikzpicture}
\end{equation}
They summarize all of the Lie algebra data of the real simple algebra.

\subsection{The Classification of Hermitian Symmetric Pairs}

Unitary discrete highest weight representations only exist for algebras $\mathfrak{g}_0$ that are part of a Hermitian symmetric pair. This is because the Cartan subalgebra should be entirely within the compact subalgebra $\mathfrak{k}_0$ (as follows from analyzing unitarity within a Cartan subgroup and the matrix realization of the Cartan involution $\theta$), which is equivalent to the Hermitian symmetric pair condition. 
Hermitian symmetric spaces are coset spaces $G/K$ (with $G$ a real group and $K$ its maximal
compact subgroup) which are Riemannian manifolds with a compatible complex structure
and on which the group $G$ acts by holomorphic
transformations. 
A manifold $X=G/K$ is Hermitian if and only if the center of $K$ is a one-dimensional central torus. They were classified by Cartan \cite{ECartan}, and fall into the list recorded in table \ref{HStable}.\footnote{Reference \cite{knapp2013lie} table (7.147).}
Crucial to us is the entry $\mathfrak{so}(2,n)$.
\begin{table}
\begin{center}
\begin{tabular}{|c|c|}
\hline 
$\mathfrak{g}_0$ & $\mathfrak{k}_0$ 
\\
\hline 
$\mathfrak{su}(p,q) $& $\mathfrak{su}(p) \oplus \mathfrak{su}(q) \oplus \mathbb{R}$
\\
\hline 
$\mathfrak{so}(2,n)$ & $\mathfrak{so}(n) \oplus \mathbb{R}$ 
\\
\hline 
$\mathfrak{sp}(n,\mathbb{R})$ & $\mathfrak{su}(n) \oplus \mathbb{R}$ 
\\
\hline $\mathfrak{so}^\ast(2n)$ & $\mathfrak{su}(n) \oplus \mathbb{R}$ 
\\
\hline E III & $\mathfrak{so}(10) \oplus \mathbb{R}$ 
\\
\hline E VII & $\mathfrak{e}_6 \oplus \mathbb{R}$ 
\\
\hline 

\end{tabular}
\end{center}
\caption{The Hermitian symmetric pairs $(\mathfrak{g}_0,\mathfrak{k}_0)$. }
\label{HStable}
\end{table}

\section{The Character Tables for Integral Unitary Weights}
\label{explicitintegralunitary}
We collect the tables of characters of integral unitary highest weight representations,
classified by their parabolic coset representative $w_i$. See section \ref{allunitaries}. Some $w_i$ are not associated with any unitary weight. In the following tables, they are signalled by an asterisk. Moreover, the brackets around $M^c_{\lambda}$ are omitted. As always, we use the notation (\ref{mathPhysTranslation}) for the weights. 
 { } \hspace{1cm}

$B_2 = \mathfrak{so}(2,3)$ 
\begin{equation}
\begin{array}{|c|c|} \hline
 w_1 & M^c_{(-E,j)} \\\hline
 w_2 & M^c_{(-E,j)}-M^c_{(-j-2,E-2)} \\\hline
 w_3 & -M^c_{(E-3,j)}+M^c_{(-E,j)}+M^c_{(-j-2,1-E)} \\\hline
 w_4 & -M^c_{(E-3,j)}+M^c_{(-E,j)}+M^c_{(-j-2,1-E)}-M^c_{(j-1,1-E)} \\\hline
\end{array}
\label{caseswiB2}
\end{equation}
To illustrate how these tables can be used, let us recover the character of the trivial representation $L_{(0,0)}$ of $\mathfrak{so}(2,3)$. This corresponds to the coset representative $w_4$, and we read in the table 
\begin{equation}
    [L_{(0,0)}] = -[M^c_{(-3,0)}]+[M^c_{(0,0)}]+[M^c_{(-2,1)}]-[M^c_{(-1,1)}] \, . 
\end{equation}
Using the explicit expression (\ref{Mcso5}), we obtain $[L_{(0,0)}]=1$, as expected.

$D_3 = \mathfrak{so}(2,4)$: 
\begin{equation}
\begin{array}{|c|c|} 
\hline
 w_1 & M^c_{(-E,j_1,j_2)} \\ \hline
 w_2 & M^c_{(-E,j_1,j_2)}-M^c_{(-j_1-3,E-3,j_2)} \\ \hline
 w_3 & M^c_{(-E,j_1,j_2)}+M^c_{(-j_1-3,-j_2-1,2-E)}-M^c_{(j_2-2,j_1,2-E)} \\ \hline
 w_4 & M^c_{(-E,j_1,j_2)}+M^c_{(-j_1-3,j_2-1,E-2)}-M^c_{(-j_2-2,j_1,E-2)} \\ \hline
 w_5 & \begin{array}{c}
       M^c_{(E-4,j_1,-j_2)}+M^c_{(-E,j_1,j_2)}-2 M^c_{(-j_1-3,1-E,-j_2)} \\
       -M^c_{(-j_2-2,j_1,E-2)}-M^c_{(j_2-2,j_1,2-E)}
 \end{array}
  \\ \hline
 w_6 & 
 \begin{array}{c}
      M^c_{(E-4,j_1,-j_2)}+M^c_{(-E,j_1,j_2)}-M^c_{(-j_1-3,1-E,-j_2)} \\
       -M^c_{(j_1-1,1-E,j_2)}+M^c_{(-j_2-2,1-E,-j_1-1)}+M^c_{(j_2-2,1-E,j_1+1)}
 \end{array}
  \\ \hline
\end{array}
\label{caseswiD3}
\end{equation}

$B_3 = \mathfrak{so}(2,5)$ 
\begin{equation}
\begin{array}{|c|c|}\hline
 w_1 & M^c_{(-E,j_1,j_2)} \\\hline
 w_2 & M^c_{(-E,j_1,j_2)}-M^c_{(-j_1-4,E-4,j_2)} \\\hline
 w_3 & M^c_{(-E,j_1,j_2)}+M^c_{(-j_1-4,j_2-1,E-3)}-M^c_{(-j_2-3,j_1,E-3)} \\\hline
 w_4 & 
  \begin{array}{c}
      -M^c_{(E-5,j_1,j_2)}+M^c_{(-E,j_1,j_2)}-M^c_{(-j_1-4,j_2-1,2-E)}\\
      +M^c_{(-j_2-3,j_1,2-E)}
 \end{array}
 \\\hline
 w_5^{\star} & 
   \begin{array}{c}
      -M^c_{(E-5,j_1,j_2)}+M^c_{(-E,j_1,j_2)}+M^c_{(-j_1-4,1-E,j_2)}\\
     +M^c_{(-j_2-3,j_1,2-E)}-M^c_{(j_2-2,j_1,2-E)}
 \end{array}
 \\\hline
 w_6 &
    \begin{array}{c}
    -M^c_{(E-5,j_1,j_2)}+M^c_{(-E,j_1,j_2)}+M^c_{(-j_1-4,1-E,j_2)}\\
     -M^c_{(j_1-1,1-E,j_2)}-M^c_{(-j_2-3,1-E,j_1+1)}+
M^c_{(j_2-2,1-E,j_1+1)}
 \end{array}
 \\\hline
\end{array}
\label{caseswiB3}
\end{equation}

$D_4 = \mathfrak{so}(2,6)$ 
\begin{equation}
    \begin{array}{|c|c|}\hline
 w_1 & M^c_{(-E,j_1,j_2,j_3)} \\\hline
 w_2 & M^c_{(-E,j_1,j_2,j_3)}-M^c_{(-j_1-5,E-5,j_2,j_3)} \\\hline
 w_3 & M^c_{(-E,j_1,j_2,j_3)}+M^c_{(-j_1-5,j_2-1,E-4,j_3)}-M^c_{(-j_2-4,j_1,E-4,j_3)} \\\hline
 w_4 & 
    \begin{array}{c}
    M^c_{(-E,j_1,j_2,j_3)}-M^c_{(-j_1-5,j_2-1,j_3-1,E-3)}+M^c_{(-j_2-4,j_1,j_3-1,E-3)}\\
     -M^c_{(-j_3-3,j_1,j_2,E-3)}
 \end{array}
 \\\hline
 w_5 & 
     \begin{array}{c}
   M^c_{(-E,j_1,j_2,j_3)}-M^c_{(-j_1-5,j_2-1,-j_3-1,3-E)}+M^c_{(-j_2-4,j_1,-j_3-1,3-E)}\\
      -M^c_{(j_3-3,j_1,j_2,3-E)}
 \end{array}
 \\\hline
 w_6 & 
     \begin{array}{c}
    M^c_{(E-6,j_1,j_2,-j_3)}+M^c_{(-E,j_1,j_2,j_3)}+2 M^c_{(-j_1-5,j_2-1,2-E,-j_3)}\\
     -2 M^c_{(-j_2-4,j_1,2-E,-j_3)}-M^c_{(-j_3-3,j_1,j_2,E-3)}-M^c_{(j_3-3,j_1,j_2,3-E)}
 \end{array}
 \\\hline
 w_7^{\star} & 
     \begin{array}{c}
   M^c_{(E-6,j_1,j_2,-j_3)}+M^c_{(-E,j_1,j_2,j_3)}-2 M^c_{(-j_1-5,1-E,j_2,-j_3)}\\
     -M^c_{(-j_2-4,j_1,2-E,-j_3)}-M^c_{(j_2-2,j_1,2-E,j_3)}+M^c_{(-j_3-3,j_1,2-E,-j_2-1)} \\
     +M^c_{(j_3-3,j_1,2-E,j_2+1)}
 \end{array}
  \\\hline
 w_8 & 
     \begin{array}{c}
     M^c_{(E-6,j_1,j_2,-j_3)}+M^c_{(-E,j_1,j_2,j_3)}-M^c_{(-j_1-5,1-E,j_2,-j_3)}\\
    -M^c_{(j_1-1,1-E,j_2,j_3)}+M^c_{(-j_2-4,1-E,j_1+1,-j_3)}+M^c_{(j_2-2,1-E,j_1+1,j_3)} \\
    -M^c_{(-j_3-3,1-E,j_1+1,-j_2-1)}-M^c_{(j_3-3,1-E,j_1+1,j_2+1)}
 \end{array}
 \\\hline
\end{array}
\label{caseswiD4}
\end{equation}

$B_4 = \mathfrak{so}(2,7)$
\begin{equation}
    \begin{array}{|c|c|} \hline
 w_1 & M^c_{(-E,j_1,j_2,j_3)} \\ \hline
 w_2 & M^c_{(-E,j_1,j_2,j_3)}-M^c_{(-j_1-6,E-6,j_2,j_3)} \\\hline
 w_3 & M^c_{(-E,j_1,j_2,j_3)}+M^c_{(-j_1-6,j_2-1,E-5,j_3)}-M^c_{(-j_2-5,j_1,E-5,j_3)} \\\hline
 w_4 & 
      \begin{array}{c}
    M^c_{(-E,j_1,j_2,j_3)}-M^c_{(-j_1-6,j_2-1,j_3-1,E-4)}+M^c_{(-j_2-5,j_1,j_3-1,E-4)}\\
    -M^c_{(-j_3-4,j_1,j_2,E-4)} 
 \end{array}
\\\hline
 w_5 & 
      \begin{array}{c}
   -M^c_{(E-7,j_1,j_2,j_3)}+M^c_{(-E,j_1,j_2,j_3)}+M^c_{(-j_1-6,j_2-1,j_3-1,3-E)}\\
  -M^c_{(-j_2-5,j_1,j_3-1,3-E)}+M^c_{(-j_3-4,j_1,j_2,3-E)} 
 \end{array}
\\\hline
 w_6^{\star} & 
      \begin{array}{c}
   -M^c_{(E-7,j_1,j_2,j_3)}+M^c_{(-E,j_1,j_2,j_3)}-M^c_{(-j_1-6,j_2-1,2-E,j_3)} \\
 +M^c_{(-j_2-5,j_1,2-E,j_3)}+M^c_{(-j_3-4,j_1,j_2,3-E)}-M^c_{(j_3-3,j_1,j_2,3-E)}
 \end{array}
  \\\hline
 w_7^{\star} & 
      \begin{array}{c}
    -M^c_{(E-7,j_1,j_2,j_3)}+M^c_{(-E,j_1,j_2,j_3)}+M^c_{(-j_1-6,1-E,j_2,j_3)}\\
  +M^c_{(-j_2-5,j_1,2-E,j_3)}-M^c_{(j_2-2,j_1,2-E,j_3)}-M^c_{(-j_3-4,j_1,2-E,j_2+1)} \\
  +M^c_{(j_3-3,j_1,2-E,j_2+1)}
 \end{array}
 \\\hline
 w_8 & 
      \begin{array}{c}
    -M^c_{(E-7,j_1,j_2,j_3)}+M^c_{(-E,j_1,j_2,j_3)}+M^c_{(-j_1-6,1-E,j_2,j_3)}\\
 -M^c_{(j_1-1,1-E,j_2,j_3)}-M^c_{(-j_2-5,1-E,j_1+1,j_3)}+M^c_{(j_2-2,1-E,j_1+1,j_3)} \\
    +M^c_{(-j_3-4,1-E,j_1+1,j_2+1)}-M^c_{(j_3-3,1-E,j_1+1,j_2+1)}
 \end{array}
 \\\hline
\end{array}
\end{equation}

\bibliographystyle{JHEP}
\bibliography{bibli.bib}

\providecommand{\href}[2]{#2}\begingroup\raggedright\endgroup

\begin{figure}[t]
    \centering
    \def\x{1.5}
    \def\y{1.5}
 \begin{tikzpicture}
\node[circle,fill=black,draw=black,inner sep=0pt,minimum size=5pt] (1) at (0.*\x,0*\y) {};\node[circle,fill=red,draw=black,inner sep=0pt,minimum size=5pt] (2) at (-1.*\x,1*\y) {};\node[circle,fill=black,draw=black,inner sep=0pt,minimum size=5pt] (3) at (0.*\x,1*\y) {};\node[circle,fill=black,draw=black,inner sep=0pt,minimum size=5pt] (4) at (1.*\x,1*\y) {};\node[circle,fill=green,draw=black,inner sep=0pt,minimum size=5pt] (5) at (-2.*\x,2*\y) {};\node[circle,fill=blue,draw=black,inner sep=0pt,minimum size=5pt] (6) at (-1.*\x,2*\y) {};\node[circle,fill=red,draw=black,inner sep=0pt,minimum size=5pt] (7) at (0.*\x,2*\y) {};\node[circle,fill=red,draw=black,inner sep=0pt,minimum size=5pt] (8) at (1.*\x,2*\y) {};\node[circle,fill=black,draw=black,inner sep=0pt,minimum size=5pt] (9) at (2.*\x,2*\y) {};\node[circle,fill=yellow,draw=black,inner sep=0pt,minimum size=5pt] (10) at (-2.5*\x,3*\y) {};\node[circle,fill=green,draw=black,inner sep=0pt,minimum size=5pt] (11) at (-1.5*\x,3*\y) {};\node[circle,fill=green,draw=black,inner sep=0pt,minimum size=5pt] (12) at (-0.5*\x,3*\y) {};\node[circle,fill=blue,draw=black,inner sep=0pt,minimum size=5pt] (13) at (0.5*\x,3*\y) {};\node[circle,fill=blue,draw=black,inner sep=0pt,minimum size=5pt] (14) at (1.5*\x,3*\y) {};\node[circle,fill=red,draw=black,inner sep=0pt,minimum size=5pt] (15) at (2.5*\x,3*\y) {};\node[circle,fill=brown,draw=black,inner sep=0pt,minimum size=5pt] (16) at (-2.*\x,4*\y) {};\node[circle,fill=yellow,draw=black,inner sep=0pt,minimum size=5pt] (17) at (-1.*\x,4*\y) {};\node[circle,fill=yellow,draw=black,inner sep=0pt,minimum size=5pt] (18) at (0.*\x,4*\y) {};\node[circle,fill=green,draw=black,inner sep=0pt,minimum size=5pt] (19) at (1.*\x,4*\y) {};\node[circle,fill=blue,draw=black,inner sep=0pt,minimum size=5pt] (20) at (2.*\x,4*\y) {};\node[circle,fill=brown,draw=black,inner sep=0pt,minimum size=5pt] (21) at (-1.*\x,5*\y) {};\node[circle,fill=brown,draw=black,inner sep=0pt,minimum size=5pt] (22) at (0.*\x,5*\y) {};\node[circle,fill=yellow,draw=black,inner sep=0pt,minimum size=5pt] (23) at (1.*\x,5*\y) {};\node[circle,fill=brown,draw=black,inner sep=0pt,minimum size=5pt] (24) at (0.*\x,6*\y) {};\draw[-] (2)-- (1);\draw[-] (3)-- (1);\draw[-] (4)-- (1);\draw[-] (5)-- (2);\draw[-] (5)-- (3);\draw[-] (6)-- (2);\draw[-] (6)-- (4);\draw[-] (7)-- (2);\draw[-] (7)-- (3);\draw[-] (8)-- (2);\draw[-] (8)-- (4);\draw[-] (9)-- (3);\draw[-] (9)-- (4);\draw[-] (10)-- (5);\draw[-] (10)-- (6);\draw[-] (10)-- (9);\draw[-] (11)-- (5);\draw[-] (11)-- (8);\draw[-] (11)-- (9);\draw[-] (12)-- (5);\draw[-] (12)-- (7);\draw[-] (13)-- (6);\draw[-] (13)-- (7);\draw[-] (13)-- (9);\draw[-] (14)-- (6);\draw[-] (14)-- (8);\draw[-] (15)-- (7);\draw[-] (15)-- (8);\draw[-] (15)-- (9);\draw[-] (16)-- (10);\draw[-] (16)-- (12);\draw[-] (16)-- (14);\draw[-] (16)-- (15);\draw[-] (17)-- (10);\draw[-] (17)-- (11);\draw[-] (17)-- (14);\draw[-] (18)-- (10);\draw[-] (18)-- (12);\draw[-] (18)-- (13);\draw[-] (19)-- (11);\draw[-] (19)-- (12);\draw[-] (19)-- (15);\draw[-] (20)-- (13);\draw[-] (20)-- (14);\draw[-] (20)-- (15);\draw[-] (21)-- (16);\draw[-] (21)-- (17);\draw[-] (21)-- (19);\draw[-] (22)-- (16);\draw[-] (22)-- (18);\draw[-] (22)-- (20);\draw[-] (23)-- (17);\draw[-] (23)-- (18);\draw[-] (23)-- (19);\draw[-] (23)-- (20);\draw[-] (24)-- (21);\draw[-] (24)-- (22);\draw[-] (24)-- (23);
\end{tikzpicture}   
\end{figure}

\end{document}